\documentclass[%
 reprint,
nofootinbib,
 amsmath,amssymb,
 aps,
prc,portrait
]{revtex4-2}
\usepackage{xcolor}
\usepackage{makecell}
\usepackage{multirow}
\usepackage{graphicx}
\usepackage{dcolumn}
\usepackage{bm}
\usepackage{rotating}
\usepackage{hyperref}
\hypersetup{colorlinks=true, citecolor=blue, urlcolor=blue, linkcolor=blue}

\begin{document}
\preprint{APS/123-QED}

\newcommand{\lanl}{Los Alamos National Laboratory, Los Alamos, NM, USA}
\newcommand{\INR}{Institute for Nuclear Research of the Russian Academy of Sciences, Moscow 117312, Russia}
\newcommand{\SNO}{SNOLAB, Sudbury, ON P3Y 1N2, Canada}
\newcommand{\Osaka}{Research Center for Nuclear Physics, Osaka University, Osaka, Japan}
\newcommand{\IfK}{Institut für Kernphysik, Westfälische Wilhelms-Universität Munster, D-48149 Munster, Germany}
\newcommand{\LBNL}{Nuclear Science Division, Lawrence Berkeley National Laboratory, Berkeley, CA 94720, USA }
\newcommand{\UCB}{Department of Physics, University of California, Berkeley, CA 94720, USA }
\newcommand{\JINR}{Joint Institute for Nuclear Research (JINR) Joliot-Curie 6, 141980, Dubna, Moscow Region, Russia}
\newcommand{\NIST}{National Institute of Standards and Technology, 100 Bureau Dr, Gaithersburg, MD 20899, USA}
\newcommand{\JSC}{JSC `State Scientific Center Research Institute of Atomic Reactors', Dimitrovgrad, 433510, Russia }
\newcommand{\UW}{Center for Experimental Nuclear Physics and Astrophysics, and Department of Physics, University of Washington, Seattle, WA 98195, USA}
\newcommand{\Carlton}{Carleton University 1125 Colonel By Drive Ottawa, K1S 5B6, Canada}
\newcommand{\UNC}{Department of Physics and Astronomy, University of North Carolina, Chapel Hill, NC 27599, USA}
\newcommand{\TUNL}{Triangle Universities Nuclear Laboratory, Durham, NC 27708, USA}

\title{A Search for Electron Neutrino Transitions to Sterile States in the BEST Experiment}

\affiliation{\INR}
\affiliation{\SNO}
\affiliation{\Osaka}
\affiliation{\lanl}
\affiliation{\IfK}
\affiliation{\LBNL}
\affiliation{\UCB}
\affiliation{\JINR}
\affiliation{\NIST}
\affiliation{\JSC}
\affiliation{\UW}
\affiliation{\Carlton}
\affiliation{\UNC}
\affiliation{\TUNL}

\author{V.V.~Barinov}\affiliation{\INR}
\author{B.T.~Cleveland}\affiliation{\SNO}
\author{S.N.~Danshin}\affiliation{\INR}
\author{H.~Ejiri}\affiliation{\Osaka}
\author{S.R.~Elliott}\affiliation{\lanl}
\author{D.~Frekers}\affiliation{\IfK}
\author{V.N.~Gavrin}\email[]{gavrin@inr.ru}\affiliation{\INR}
\author{V.V.~Gorbachev}\affiliation{\INR}
\author{D.S.~Gorbunov}\affiliation{\INR}
\author{W.C.~Haxton}\affiliation{\LBNL}\affiliation{\UCB}
\author{T.V.~Ibragimova}\affiliation{\INR}
\author{I.~Kim}\affiliation{\lanl}
\author{Yu.P.~Kozlova}\affiliation{\INR}
\author{L.V.~Kravchuk}\affiliation{\INR}
\author{V.V.~Kuzminov}\affiliation{\INR}
\author{B.K.~Lubsandorzhiev}\affiliation{\INR}
\author{Yu.M.~Malyshkin}\affiliation{\INR}
\author{R.~Massarczyk}\affiliation{\lanl}
\author{V.A.~Matveev}\affiliation{\JINR}
\author{I.N.~Mirmov}\affiliation{\INR}
\author{J.S.~Nico}\affiliation{\NIST}
\author{A.L.~Petelin}\affiliation{\JSC}
\author{R.G.H.~Robertson}\affiliation{\UW}
\author{D.~Sinclair}\affiliation{\Carlton}
\author{A.A.~Shikhin}\affiliation{\INR}
\author{V.A.~Tarasov }\affiliation{\JSC}
\author{G.V.~Trubnikov}\affiliation{\JINR}
\author{E.P.~Veretenkin}\affiliation{\INR}
\author{J.F.~Wilkerson}\affiliation{\UNC}\affiliation{\TUNL}
\author{A.I.~Zvir}\affiliation{\JSC}

\date{\today}

\begin{abstract}
The Baksan Experiment on Sterile Transitions~(BEST) probes the gallium anomaly and its possible connections to oscillations between active and sterile neutrinos. 
Based on the Gallium-Germanium Neutrino Telescope~(GGNT) technology of the SAGE experiment, BEST employs two zones of liquid Ga target to explore neutrino oscillations on the meter scale. 
Oscillations on this short scale could produce deficits in the $^{71}$Ge production rates within the two zones, as well as a possible rate difference between the zones. 

From July 5th to October 13th 2019, the two-zone target was exposed to a primarily monoenergetic, 3.4-MCi $^{51}$Cr neutrino source 10 times for a total of 20 independent $^{71}$Ge extractions from the two Ga targets. 
The $^{71}$Ge production rates from the neutrino source were measured from July 2019 to March 2020. 
At the end of these measurements, the counters were filled with $^{71}$Ge doped gas and calibrated during November 2020.
In this paper, results from the BEST sterile neutrino oscillation experiment are presented in details.
The ratio of the measured $^{71}$Ge production rates to the predicted rates for the inner and the outer target volumes are calculated from the known neutrino capture cross section. 
Comparable deficits in the measured ratios relative to predicted values are found for both zones, with the $4 \sigma$ deviations from unity consistent with the previously reported gallium anomaly. 
If interpreted in the context of neutrino oscillations, the deficits give best fit oscillation parameters of $\Delta m^2=3.3^{+\infty}_{-2.3}$~eV$^2$ and sin$^2 2\theta=0.42^{+0.15}_{-0.17}$, consistent with $\nu_e \rightarrow \nu_s$ oscillations governed by a surprisingly large mixing angle.

\end{abstract}

\maketitle

\section{Introduction} \label{sec:introduction}

\begin{figure*}[thp]
 \centering
 \includegraphics[width=0.8\textwidth]{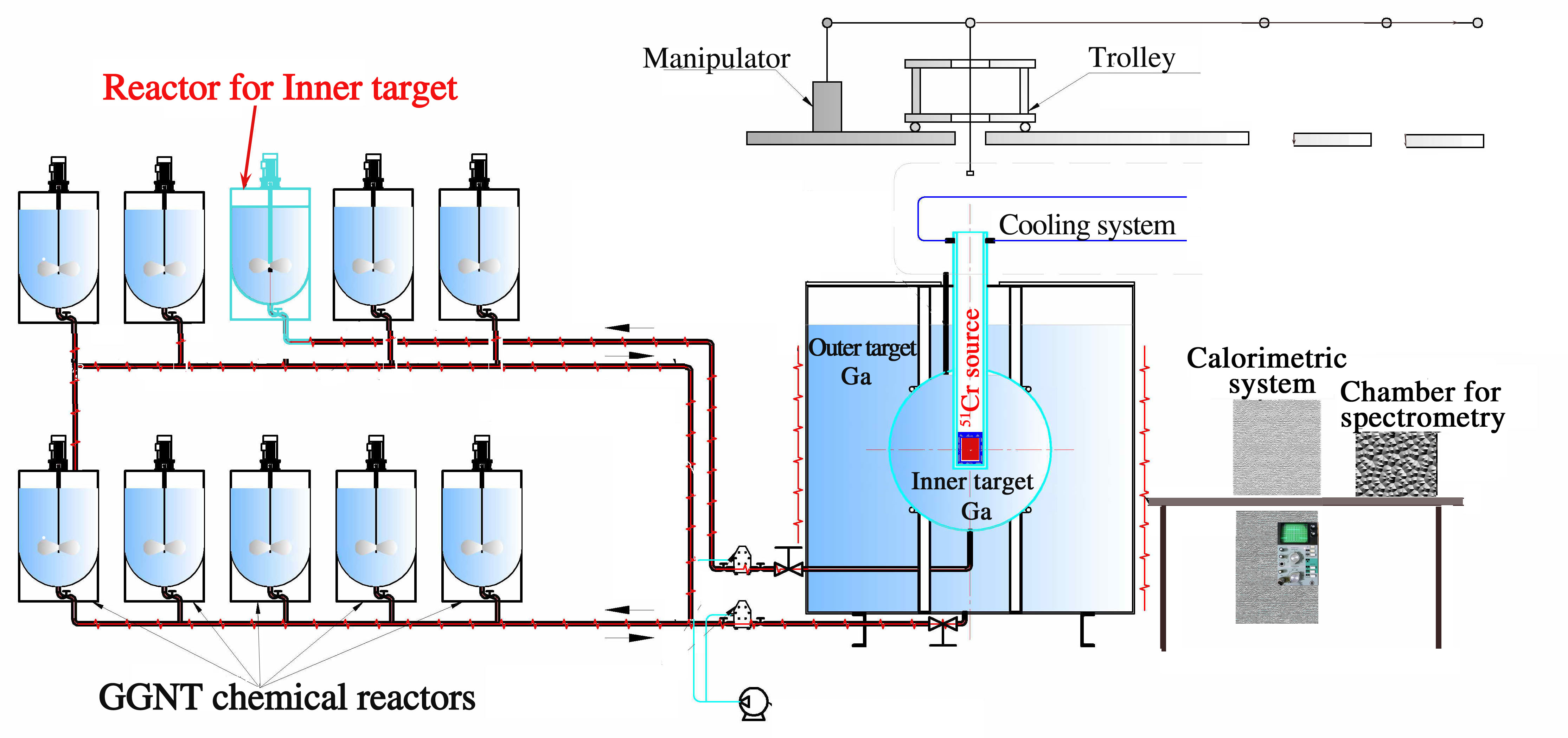}
 \caption{The Ga target and extraction piping diagram also indicating the source handling apparatus.}
 \label{fig:best_scheme}
\end{figure*}

Sterile neutrinos~($\nu_s$) are hypothetical fermions which are singlets with respect to the Standard Model~(SM) gauge groups~\cite{pontecorvo1968neutrino,gariazzo2015light,giunti2019annual,boser2020status,diaz2020where,seo2021review,dasgupta2021sterile}. 
Lacking electric, color, or hypercharges, they would not participate in electromagnetic, strong or weak interactions, but could thorough their gravitational interactions have consequences for astrophysics. 
In particular, sterile neutrinos could play a role in the formation of large-scale structure while evading the strong three-flavor constraints imposed from multiple experiments~\cite{janot2019improved,DELPHI2003photon,L32003single,aleph2006precision,Planck2018results,Fields2019bigbang,Ivanov2019cosmological}. 
Sterile neutrinos naturally arise in many extensions of the Standard Model and have been invoked to account for various anomalies, including~(for specific parameters) dark matter.

The existence of singlet state neutrinos is theoretically well-motivated. 
Neutrinos are the only fermions known to have an intrinsic left-handed chirality. 
When their right-handed counterparts are considered, the simplest neutrino mass generation model of the type-I seesaw mechanism is allowed~\cite{glashow1980future,schechter1980neutrino,mohapatra1980neutrino,minkowski1977rate}. 
The inactive right-handed neutrinos have so-called Majorana masses independent of the Higgs mechanism and can generate neutrino masses at any scale. 
Assuming the existence of sterile neutrinos can therefore not only explain the non-zero mass of neutrinos but also account for neutrino masses being at least 5 orders of magnitude smaller than the electron rest mass.

All plausible couplings of sterile neutrinos are weak and are thought to be beyond the detection capability of the current detectors. 
Hence, the only possibility for detecting sterile neutrinos is to observe the mixing with active neutrinos.
In general, extensions of the SM to include sterile neutrinos do not impose strong constraints on their masses. 
While sterile neutrinos have been proposed with masses ranging from sub-eV to GUT scales,  the 1~eV scale is particularly interesting as multiple physical phenomena can be explained by a single hypothesis. 
Such indications include unexplained excesses in the electron neutrino fluxes in LSND~\cite{LSND1996evidence,LSND2001evidence} and MiniBooNE experiments; reactor antineutrino anomalies at 10-100~m oscillation lengths~\cite{mueller2011improved,chooz2011reactor,adamson2016reactor}; and the anomalous deficits in the radiochemical-source measurements of the SAGE~\cite{sage1999source,sage2006argon} and the GALLEX~\cite{gallex1998source,gallex2010reanalysis}  experiments.
Recent results from MicroBooNE indicate a discrepancy with the MiniBooNE result~\cite{abratenko2021search}, but do not rule out the sterile neutrino hypothesis~\cite{denton2021sterile,arguelles2021microboone}.

\begin{figure*}[thp]
 \centering
 \includegraphics[width=0.8\textwidth]{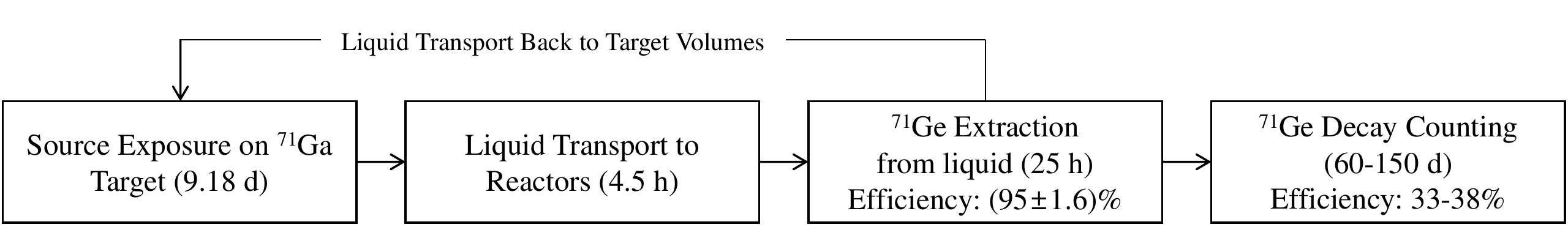}
 \caption{Flowchart showing the experimental steps of the BEST experiment. Time spent on each step and relevant efficiencies are also presented. }
 \label{fig:best_flowchart}
\end{figure*}

The Baksan Experiment on Sterile Transitions (BEST) is a two-distance oscillation experiment designed to explore the deficit of electron neutrinos~($\nu_e$) previously reported in the SAGE and GALLEX source experiments, commonly known as the gallium anomaly. 
The experimental design is presented in Fig.~\ref{fig:best_scheme}, with the flowchart showing the experimental steps in Fig.~\ref{fig:best_flowchart}.
About 47 tonnes of liquid Ga metal, divided into two concentric zones, serve as the target for the charged-current reaction $^{71}$Ga$(\nu,e)$ $^{71}$Ge.
The decay of  the mainly monoenergetic neutrino source of the $^{51}$Cr isotope is via electron capture with the emission of $\nu_e$ in four spectral lines.
The $\nu_e$ energies and their branching ratios are: 747~keV~(81.63\%), 427~keV~(8.95\%), 752~keV~(8.49\%) and 432~keV~(0.93\%)~\cite{acero2008limits}.
The source was placed at the center of the two zones of the Ga target, irradiating the two volumes simultaneously. 
This allows the production rates of $^{71}$Ge to be measured at two different distances.
As the sizes of the two target zones are roughly at a meter scale, BEST has a high sensitivity to the oscillatory behaviour in the $\nu_e$ flux at the same scale, which corresponds to $\Delta m^2 \approx 1$~eV$^2$. 
Neutrino oscillations at this short scale would be indicated by deficits in the $^{71}$Ge production rates within the two zones, as well as the differences between them. 
The use of an electron capture $\nu_e$ source is a powerful technique to search for neutrino oscillations. The $\nu_e$ spectrum is simple, being comprised of a dominant monoenergetic component and 3 sub-dominant lines. It is a well-understood spectrum relying on well-known nuclear and atomic physics parameters.

In this article, we  present the results from the BEST $\nu_s$ oscillation search experiment. 
Section~\ref{sec:exposure} describes the geometry of the two-zone gallium target which is irradiated by a $^{51}$Cr $\nu_e$ source. 
Section~\ref{sec:extraction} reviews how the $^{71}$Ge atoms are extracted from the target volumes and transferred to the proportional counters. 
Section~\ref{sec:counting} is devoted to a description of the counter systems which measure the $^{71}$Ge decay rates.
In Section~\ref{sec:analysis}, techniques of candidate event selection and analysis to set the exclusion within the $(\Delta m^2, \textrm{sin}^2(2\theta))$ oscillation parameter space are introduced. 
We discuss the first result from the BEST experiment in Section~\ref{sec:discussion}, and conclude in Section~\ref{sec:conclusion} that our result is consistent with the neutrino oscillation hypothesis with $\Delta m^2>1$~eV$^2$ sterile neutrinos. 

\section{Target Geometry} \label{sec:exposure}

The Baksan Neutrino Observatory~(BNO) is located in a dedicated underground laboratory in the northern Caucasus mountains of Russia. 
Located 3.5~km below the surface, the deep underground laboratory has an overburden of 4700~meters-water-equivalent~(mwe), resulting in muon flux in the laboratory of $(3.03\pm0.10)\times10^{-9}$/(cm$^2$ s)~\cite{sage2006argon}.
The entirety of the laboratory is lined with 60~cm of radiopure concrete and 6~mm steel shell to reduce $\gamma$ and neutron backgrounds from rocks. 

The concentric two-zone gallium target of the BEST experiment is located in the main hall of the laboratory. 
The inner spherical volume with inner diameter 133.5~cm contains (7.4691$\pm$0.0631)~t of liquid Ga, and the outer cylindrical volume with  inner diameter 218~cm and maximum Ga target height 211.2~cm contains (39.9593$\pm$0.0024)~t of Ga. 
The Ga metal, heated to 30.0~\textdegree C, remains molten inside the volumes. 
The carrier Ge and any produced $^{71}$Ge was extracted at the end of each exposure for the calculation of production rates.
The extraction procedures are explained in Sec.\ref{sec:extraction_efficiency}, and discussed more thoroughly in Ref.~\cite{sage1999solar}.

The $^{51}$Cr source with known activity was contained in a cylindrical cartridge with a radius of 4.3~cm and height of 10.8~cm. 
This source assembly was placed in a specially constructed tungsten radiation shield with a thickness of $\approx$30~mm and a weight of 42.8~kg, which provides radiation safety from gamma activity from the source, and a steel shell with a special cap for capturing the source by a manipulator.
The assembly was placed at the center of the two zones to irradiate both volumes simultaneously. 

The source was delivered to BNO on July 5, 2019 and was placed into the two-zone target at 14:02 local time~(UTC+3) that same day, and this is our chosen reference time for the source strength. 
The source was manufactured by irradiating 4~kg of $^{50}$Cr-enriched metal for 100~d in a reactor at the State Scientific Center Research Institute of Atomic Reactors, Dimitrovgrad, Russia. 
The activity~($A$) of the source is measured by calorimetric method based on the heat released by the source~\cite{kozlova2020measurement,Gavrin_2021}. 
At the end of each extraction, the source is moved into a lead container for the activity measurement. The $\gamma$ spectrum is measured at 21.65~m distance with a Ge detector for an hour. 
After the spectral measurement, the source is moved to the calorimeter to measure the heat emitted by the source for 20-21 hours. 
The tungsten shield of the source does not affect calorimetry as the shield is included within the calorimeter. Moreover, it helps by absorbing gamma radiation from impurities in the source, increasing the accuracy of determining the heat release from $^{51}$Cr.
Ten calorimetric measurements were performed after nine days of irradiation of gallium targets.
The decay of $^{51}$Cr ($\approx$90\% to the ground state and $\approx$10\% to the excited state) releases on average ($36.750\pm0.084$)~keV per decay event.
From spectrometric measurements of gamma radiation of the source, the heat contribution from radioactive impurities is found to be ($3.7\pm 0.5$)~mW, which is $5\times10^{-6}$ of the initial $^{51}$Cr source power. 
The measured activity of the source at the reference time is (3.414$\pm$0.008)~MCi. 
The measured half-life of the $^{51}$Cr source is (27.710$\pm$0.017)~days, which is in a good agreement with the known $^{51}$Cr half-life of (27.704$\pm$0.017)~days~\cite{Zhou1991}.
A full description of the source and the calorimetric measurements of its intensity can be found in Refs.~\cite{kozlova2020measurement,Gavrin_2021}.

\section{$^{71}$G\lowercase{e} Extraction} \label{sec:extraction}

\begin{table*}[thp]
    \caption{Extraction and exposure details for the inner target. The times of exposure are given in days of year 2019. A mass of (7.4691$\pm$0.0631)~t of liquid Ga was irradiated by the neutrino source in the inner target.}
    \centering
    \footnotesize
    \begin{tabular}{|c|c||c|c||c|c|c||c|c|}
    \hline
        \multicolumn{2}{|c||}{Extraction} & \multicolumn{2}{c||}{Source exposure} & \multicolumn{3}{c||}{Solar exposure} 	& \multicolumn{2}{c|}{Extraction efficiency} \\
        \hline
        \multirow{2}{*}{Name} &\multirow{2}{*}{Date (2019)} & Begin & End    &  Begin & End	& Ga mass 	& \multirow{2}{*}{from Ga} & \multirow{2}{*}{into GeH$_4$}  \\
             & 	             & 	(day in year)	& (day in year) & 	(day in year)	& (day in year)	& (tonnes)	&	&  \\
        \hline
        Inner-1	& 15 Jul 16:01	& 186.585	& 196.376	& 183.667	& 196.667	& 7.594	& 0.9747(97)	& 0.9460(123)\\
        Inner-2	& 25 Jul 16:32	& 197.362	& 206.372	& 196.689	& 206.689	& 7.586	& 0.9814(98)	& 0.9559(124)\\
        Inner-3	& 04 Aug 16:37	& 207.282	& 216.374	& 206.792	& 216.692	& 7.578	& 0.9795(98)	& 0.9673(126)\\
        Inner-4	& 14 Aug 15:35	& 217.286	& 226.371	& 216.749	& 226.649	& 7.57	& 0.9801(98)	& 0.9515(124)\\
        Inner-5	& 24 Aug 17:17	& 227.258	& 236.458	& 226.620	& 236.720	& 7.562	& 0.9808(98)	& 0.9554(124)\\
        Inner-6	& 03 Sep 15:18	& 237.342	& 246.369	& 236.738	& 246.638	& 7.554	& 0.9818(98)	& 0.9548(124)\\
        Inner-7	& 13 Sep 15:11	& 247.243	& 256.368	& 246.733	& 246.733	& 7.546	& 0.9813(98)	& 0.9381(122)\\
        Inner-8	& 23 Sep 15:17	& 257.241	& 266.369	& 256.737	& 266.637	& 7.538	& 0.9835(98)	& 0.9789(127)\\
        Inner-9	& 03 Oct 15:00	& 267.24	& 276.369	& 266.725	& 276.625	& 7.529	& 0.9824(98)	& 0.9545(123)\\
        Inner-10	& 13 Oct 14:59	& 277.201	& 286.367	& 276.724	& 286.624	& 7.521	& 0.9806(98)	& 0.9372(122)\\
        \hline
    \end{tabular}
    \label{tab:inner_extraction}
\end{table*}

\begin{table*}[thp]
    \caption{Extraction and exposure details for the outer target. The times of exposure are given in days of year 2019. A mass of (39.9593$\pm$0.0024)~t of liquid Ga was irradiated by the neutrino source in the outer target.}
    \centering
    \footnotesize
    \begin{tabular}{|c|c||c|c||c|c|c||c|c|}
    \hline
        \multicolumn{2}{|c||}{Extraction} & \multicolumn{2}{c||}{Source exposure} & \multicolumn{3}{c||}{Solar exposure} 	& \multicolumn{2}{c|}{Extraction efficiency} \\
        \hline
        \multirow{2}{*}{Name} &\multirow{2}{*}{Date (2019)} & Begin & End &  Begin & End	& Ga mass 	& \multirow{2}{*}{from Ga} & \multirow{2}{*}{into GeH$_4$}  \\
             & 	             & 	(day in year)	& (day in year) & 	(day in year)	& (day in year)	& (tonnes)	&	& 	\\
        \hline
        Outer-1	& 15 Jul 13:59	& 186.585	& 196.376	& 183.783	& 196.583	& 44.237	& 0.9868(99)	& 0.9503(124)\\
        Outer-2	& 25 Jul 13:51	& 197.362	& 206.372	& 196.877	& 206.577	& 44.191	& 0.9841(98)	& 0.9581(125)\\
        Outer-3	& 04 Aug 12:47	& 207.282	& 216.374	& 206.833	& 216.533	& 44.145	& 0.9881(99)	& 0.9668(126)\\
        Outer-4	& 14 Aug 12:51	& 217.286	& 226.371	& 216.835	& 226.535	& 44.098	& 0.9858(99)	& 0.9622(125)\\
        Outer-5	& 24 Aug 14:35	& 227.258	& 236.458	& 226.808	& 236.608	& 44.052	& 0.9871(99)	& 0.9609(125)\\
        Outer-6	& 03 Sep 12:35	& 237.342	& 246.369	& 236.924	& 246.524	& 44.004	& 0.9893(99)	& 0.9253(120)\\
        Outer-7	& 13 Sep 12:29	& 247.243	& 256.368	& 246.82	& 256.520	& 43.954	& 0.9904(99)	& 0.9514(124)\\
        Outer-8	& 23 Sep 12:32	& 257.241	& 266.369	& 256.822	& 266.522	& 43.906	& 0.9897(99)	& 0.9897(129)\\
        Outer-9	& 03 Oct 12:27	& 267.24	& 276.369	& 266.819	& 276.519	& 43.857	& 0.9881(99)	& 0.9664(126)\\
        Outer-10& 13 Oct 12:26	& 277.201	& 286.367	& 276.818	& 286.518	& 43.807	& 0.9877(99)	& 0.9538(124)\\
        \hline
    \end{tabular}
    \label{tab:outer_extraction}
\end{table*}

Between July~5th and October~13th 2019, the two-zone target was exposed to the source 10 times with an average exposure time of 9.18~d. 
The irradiation schedule was devised to maximize the number of extracted $^{71}$Ge atoms. 
The extraction was carried out at the end of each exposure.
The first extraction counting periods were shorter in time due to the limited number of available counters. 
The shorter counting time has almost no effect on the number of calculated $^{71}$Ge decays but increases the statistical uncertainty due to the lower statistics of the measured counter background. 
After each exposure period, the Ga targets were transferred to chemical reactor vessels, and the produced $^{71}$Ge atoms were extracted from Ga with the technique almost identical to the one used for the SAGE experiment~\cite{sage1999solar,sage2006argon}. 
Here we provide an overview. 

The Ge carrier and the $^{71}$Ge atoms produced by the neutrino capture reaction are extracted from the Ga targets into an aqueous solution by adding a weak acidic solution of H$_2$O$_2$~\cite{bahcall1978proposed,barabanov1985pilot}, ensuring independent extractions of $^{71}$Ge atoms from each zone of the Ga target.
The acid oxidizes the Ge metal and GeH$_4$ is synthesized in gaseous form. 
Gallium is pumped out of the vessels until the aqueous acidic solution on the surface of gallium begins to be pumped out.
The GeH$_4$ is then mixed with Xe gas and introduced into a low-background proportional counter. 

The extraction efficiency of Ge from a large mass of Ga is determined with high accuracy using mass spectrometry~\cite{CLEVELAND201541}. 
For BEST, the process of the efficiency estimation follows the same procedure as was used in SAGE. 
A known amount of stable $^{72}$Ge (2.4~$\mu$mol, 92\%) and $^{76}$Ge~(2.4~$\mu$mol, 95\%) carriers were added to the volume in form of Ga-Ge alloy as the stable Ge carrier, and the molten Ga was stirred thoroughly to disperse the Ge throughout the volumes~\cite{sage1999solar}.  
The total extraction efficiency is given by the ratio of the amount of Ge in the synthesized GeH$_4$ to the initial amount of Ge in the Ga targets at the beginning of the exposure.
Taking into account all the factors that can affect the extraction efficiency including the temperature, the amount of oxidizing agent~(H$_2$O$_2$), and the volume of the aqueous phase which defines the time of later concentration of Ge, the extraction procedure reached (98$\pm$0.2)\% efficiency from Ga, and the total efficiency including synthesis reached ($95\pm1.6$)~\%.
The extraction and exposure details for the inner and outer targets are summarized in Tables~\ref{tab:inner_extraction} and~\ref{tab:outer_extraction}, respectively.
Due to the inefficiency of the extraction, a small proportion of the carrier Ge is still present within the Ga. The numbers of these carryover atoms were estimated and accounted for in the production rate calculation.

At each source exposure, gallium was pumped with the same levels into the two volumes, which were monitored by level sensors in the zones.
Therefore, a larger mass of gallium was irradiated in solar neutrino exposures than in exposures with a source.
This contribution was also accounted for.

\section{Counting of $^{71}$G\lowercase{e}} \label{sec:counting}

\begin{table*}[t]
    \caption{Counting parameters for the inner target. $\Delta$ is the exponentially weighted live time. The live time and $\Delta$ include all time cuts. The counter efficiency for each extraction has the same fractional uncertainty of $-2.1/+2.3$\%, as explained in details in Sec.~\ref{sec:syst_uncer}.}
    \centering
    \small
    \begin{tabular}{|c|c||c|c||c|c||c||c|c|c|c|}
    \hline
        \multicolumn{2}{|c||}{Extraction} & \multicolumn{2}{c||}{Counter filling} & \multicolumn{2}{c||}{\makecell{Counter efficiency \\ (after event selection)}} 	& \multirow{2}{*}{\makecell{Day counting \\ began \\ (day in year)}} &  \multicolumn{2}{c|}{Live time (days)} &   \multicolumn{2}{c|}{$\Delta$} \\
        \cline{1-6}
        \cline{8-11}
        Run & \makecell{Counter \\ name} & \makecell{Pressure \\ (mmHg) } & \makecell{GeH$_4$ \\ fraction \\ (\%)} & 
        K-peak & L-peak &  & K-peak & L-peak & K-peak & L-peak
        \\
        \hline
        Inner-1	& YCT92	& 630	& 8.8	& 0.3663	& 0.3803	& 197.66	& 54.478	& 34.364	& 0.8102	& 0.7450 \\
        Inner-2	& YCT2	& 640	& 9.5	& 0.3647	& 0.3785	& 207.623	& 53.706	& 29.834	& 0.7839	& 0.6542 \\
        Inner-3	& YCN43	& 650	& 9.3	& 0.3605	& 0.3599	& 217.693	& 50.525	& 50.525	& 0.7143	& 0.7143 \\
        Inner-4	& YCT97	& 640	& 9.2	& 0.3679	& 0.3769	& 227.644	& 52.808	& 29.884	& 0.7872	& 0.3672 \\
        Inner-5	& YCN46	& 650	& 9.5	& 0.3649	& 0.3654	& 237.790	& 150.436	& 150.436	& 0.7470	& 0.7470 \\
        Inner-6	& YCN42	& 640	& 9.8	& 0.3577	& 0.3604	& 247.597	& 140.143	& 133.113	& 0.7717	& 0.3892 \\
        Inner-7	& YCT92	& 640	& 9.3	& 0.3676	& 0.3793	& 257.617	& 129.483	& 130.843	& 0.7493	& 0.6776 \\
        Inner-8	& YCT2	& 645	& 9.5	& 0.3656	& 0.3779	& 267.634	& 129.060	& 131.764	& 0.7754	& 0.7855 \\
        Inner-9	& YCN43	& 640	& 9.1	& 0.359	    & 0.3610	& 277.678	& 152.034	& 152.034	& 0.8019	& 0.8019 \\
        Inner-10	& YCT97	& 650	& 9.1	& 0.3698	& 0.3755	& 287.625	& 144.446	& 147.014	& 0.7629	& 0.7955 \\
        \hline
    \end{tabular}
    \label{tab:inner_counting}
\end{table*}

\begin{table*}[t]
    \caption{Counting parameters for the outer target. $\Delta$ is the exponentially weighted live time. The live time and $\Delta$ include all time cuts. The counter efficiency for each extraction has the same fractional uncertainty of $-2.1/+2.3$\%, as explained in details in Sec.~\ref{sec:syst_uncer}.}
    \centering
    \small
    \begin{tabular}{|c|c||c|c||c|c||c||c|c|c|c|}
    \hline
        \multicolumn{2}{|c||}{Extraction} & \multicolumn{2}{c||}{Counter filling} & \multicolumn{2}{c||}{\makecell{Counter efficiency \\ (after event selection)}} 	& \multirow{2}{*}{\makecell{Day counting \\ began \\ (day in Year)}} &  \multicolumn{2}{c|}{Live time (days)} &   \multicolumn{2}{c|}{$\Delta$} \\
        \cline{1-6}
        \cline{8-11}
        Run & \makecell{Counter \\ Name} & \makecell{Pressure \\ (mmHg) } & \makecell{GeH$_4$ \\ fraction \\ (\%)} & 
        K-peak & L-peak &  & K-peak & L-peak & K-peak & L-peak
        \\
        \hline
        Outer-1	& YCN113& 635	& 9.5   & 0.3422	& 0.3596	& 197.66	& 53.788	& 33.662	& 0.7648	& 0.6996\\
        Outer-2	& YCT3	& 635	& 9.5	& 0.3707	& 0.3792	& 207.623	& 54.376	& 30.640	& 0.8043	& 0.6755\\
        Outer-3	& YCNA9	& 640	& 10.5	& 0.2933	& 0.3358	& 217.693	& 51.070	& 51.070	& 0.7650	& 0.7650\\
        Outer-4	& YCT9	& 635	& 9.6	& 0.3658	& 0.381	   &  227.644	& 52.981	& 30.423	& 0.7820	& 0.3755\\
        Outer-5	& YCN41	& 635	& 10.0	& 0.3568	& 0.3727	& 237.790	& 147.774	& 147.774	& 0.8025	& 0.8025\\
        Outer-6	& YCT4	& 630	& 9.0	& 0.3585	& 0.3577	& 247.597	& 139.382	& 131.148	& 0.8012	& 0.3843\\
        Outer-7	& YCN113& 630	& 10.3	& 0.3407	& 0.3607	& 257.617	& 134.985	& 136.161	& 0.7977	& 0.7108 \\
        Outer-8	& YCT3	& 640	& 9.5	& 0.3716	& 0.3785	& 267.634	& 129.098	& 131.802	& 0.8298	& 0.8398 \\
        Outer-9	& YCNA9	& 635	& 9.9	& 0.293	    &  0.3360	& 277.678	& 155.439	& 155.439	& 0.7865	& 0.7865 \\
        Outer-10	& YCT9	& 645	& 9.5	& 0.3677	& 0.3797	& 287.625	& 143.604	& 146.307	& 0.7567	& 0.7905 \\
        \hline
    \end{tabular}
    \label{tab:outer_counting}
\end{table*}

The decay process $^{71}$Ge + $e^-$ $\rightarrow ^{71}$Ga +$\nu$ has a half-life of 11.4~d~\cite{hampel1985}. 
Two peaks at 10.4~keV~(K peak) and 1.2~keV~(L peak) are observable in the proportional counters. 
The K-capture, which constitutes $\approx$88\% of all decays, can release 10.4~keV Auger electrons~(41.5\%), 1.2~keV Auger electrons with 9.2~keV X~rays~(41.2\%), or 0.1~keV Auger electrons with 10.3~keV X~rays~(5.3\%)~\cite{sage1999solar}. 
On the other hand, L and M captures give almost entirely Auger electrons of energies 1.2~keV and 0.12~keV respectively~\cite{sage1999solar}. 
Since the proportional counters have much higher efficiency for Auger electrons than for the X~rays, the number of K and L peak counts are almost equal.
The pulse shapes recorded from the counters are analyzed to suppress the contribution from unwanted backgrounds.

The design of the proportional counters used is described in detail in Refs.~\cite{danshin1994proportional,sage1999solar,abdurashitov2009measurement}. 
The proportional counters, filled with GeH$_4$ from each extraction, are placed in the well of a NaI veto detector within a large passive shield. 
To suppress the $^{222}$Rn background, the shield volume is purged with boil-off gas from the liquid nitrogen. 
The average counting time for the first four runs is approximately 50~d and was extended to about 140~d in the later six runs to account for the decreased source strength.
Calibration data with an $^{55}$Fe source were taken once every 2 weeks throughout the measurements to ensure the stability of counting. 
Mean variations of the positions of the calibration peak during full counting period were 1.4\% and 1.3\% for the inner and the outer volumes and taken into account in analysis.
Two periods during the counting were excluded from the analysis due to identified issues:

\begin{enumerate}
    \item The time response of all slots in high-gain channel (L-peak) was slow between 2019-08-23 07:03 and 2019-09-14 14:49.
    \item There was a failure in the low gain channel (K-peak) between 2019-11-06 21:10 and 2019-11-09 20:45.
\end{enumerate}

\noindent
The livetime of some runs were affected accordingly. 

After the counting of the extraction samples from the Cr experiment was completed in the spring-fall of 2020, the counting efficiency was directly measured for each counter.
The volume efficiencies, accounting for the dead volume near the cathode, were directly measured with $^{37}$Ar. 
The event selection efficiencies of the waveform analysis were measured with $^{71}$Ge calibrations. 
The rise-time~($T_N$) values from the $^{71}$Ge calibration data were arranged in ascending order and an upper limit set such that 4\% of the calibration events were excluded~\cite{sage2006argon}. 
Details of the event selection and the $^{71}$Ge calibration runs to verify the selection scheme are summarized in Sec.~\ref{sec:event_selection} and Appx.~\ref{app:ge71} respectively. 

The calculated counting efficiency using the measured pressure in the counter, GeH$_4$ fraction, and $^{37}$Ar volume efficiency was determined for each extraction. They are summarized in Tables~\ref{tab:inner_counting} and~\ref{tab:outer_counting} with other counting parameters. 
Values of the exponentially weighted livetime $\Delta$ are also presented for each extraction

\begin{equation}
    \Delta = \sum_{k=1}^n (e^{-\lambda t_{bk}} -e^{-\lambda t_{ek}})~,
\end{equation}

\noindent
where the sum is over $n$ counting intervals, each of which has a starting time $t_{bk}$ and an ending time $t_{ek}$. $\lambda$ is the decay constant of $^{71}$Ge.

\section{Analysis} \label{sec:analysis}

A primary analysis on the BEST data was performed by the analysis group in the Institute for Nuclear Research, Russia, and was verified by an independent analysis carried out by a separate analysis group in Los Alamos National Laboratory. 
Both of the analyses obtained similar results to within about 2\%. This difference is due to minor event-selection differences at the edges of the selection borders in energy and rise time. 
This difference is accounted for by the estimated systematic uncertainties in the efficiencies for those cuts. 

Pulses from the proportional counters are digitized at two different gains. 
The higher gain channel is chosen for the $^{71}$Ge L-peak, and the lower gain channel is used for the K-peak analysis.
The pulse shape analysis is performed to differentiate the $^{71}$Ge signals from backgrounds. 

The analytic form of pulse shapes used in the BEST analysis is derived from the model studied in Ref.~\cite{elliott1990analytical} where

\begingroup
\footnotesize
\begin{equation}\label{eqn:pulseModel}
    \begin{split}
        V&(0<t<T_N)=V_0\Big[\frac{t+t_0}{T_N} \mathrm{ln}\Big(1+\frac{t}{t_0}\Big)-\frac{t}{T_N}\Big]~,\\
        V&(t>T_N)=V_0\Big[\mathrm{ln}\Big(1+\frac{t-T_N}{t_0}\Big) -1 - \frac{t+t_0}{T_N}\mathrm{ln}\Big(1-\frac{T_N}{t+t_0}\Big)\Big]~,
    \end{split}
\end{equation}
\endgroup

\noindent
with $V(t<0)=0$. 
Here, $T_N$ is the rise-time, which is the duration over which the ionization arrives at the anode, $t_0$ is the time inversely proportional to the ion mobility, and $V_0$ is the voltage proportional to the total amount of ionization.
We fit every pulse that is not identified as saturation or breakdown to Eq.~\ref{eqn:pulseModel}. The fit is made from 40~ns before the time of pulse onset to 400~ns after onset. 
Five parameters are determined by the fit: $t_\textrm{onset}, V_\textrm{onset}, V_0, t_0$, and $T_N$.
The integral of the fit function over pulse waveform for 800~ns after the onset is the energy parameter~\cite{sage1999source}.
The $T_N$ value is used as a background rejection parameter. 

\subsection{Calibration}\label{sec:calibration}

Following the procedure described in Ref.~\cite{sage1999source}, we first find the position and the width of the 5.895~keV calibration peak in each calibration run. 
The expected locations for the $^{71}$Ge K and L peaks are then calculated from the energy ratios with corrections for the non-linearity factors as described in Ref.~\cite{sage1999solar}.
For the counters with the GeH$_4$ fraction $G$, counter pressure $P$, and operating voltage $V$, the non-linearity factors for the location $P_K$ and the resolution $R_K$ of the K peak follow the empirical formulas reported in Ref.~\cite{sage1999solar} as

\begin{equation}
    \footnotesize
    \begin{split}
    \frac{P_K(^{71}\mathrm{Ge})}{P(^{55}\mathrm{Fe})} &= \frac{10.367}{5.895}[1-(4.5G+2.78)(V-V_\mathrm{crit1})\times10^{-6}]~, \\
    \frac{R_K(^{71}\mathrm{Ge})}{R(^{55}\mathrm{Fe})} &= \sqrt{\frac{5.895}{10.367}}[1+1.5\times10^{-3}(V-V_\mathrm{crit2})]~,
    \end{split}
\end{equation}

\noindent
where the corrections are applied if $V$ is above the respective critical voltages $V_\mathrm{crit1} = 10.5G+0.6P+588$ and $V_\mathrm{crit2}=6G+P/3+824$. 
The typical corrections for the K-peak position and the resolution are 2\% and 15\%, respectively, and no corrections are required for the L peak as the critical voltages are much higher than for the K peak. 

\begin{table}[!t]
    \caption{The cut values on $T_N$ of the pulses for $^{71}$Ge K and L peak for each counting in the inner target. Events with $T_N$ values greater than each cut value are selected out.}
    \centering
    \footnotesize
    \begin{tabular}{|c|c|c|c|c|c|c|}
    \hline
    \multicolumn{4}{|c|}{Counter Filling} &  \multicolumn{2}{c|}{$T_N$}\\
    \hline
    \makecell{Extraction \\name}	& \makecell{Counter \\name}	& \makecell{Pressure \\(mmHg)}	& \makecell{GeH$_4$ \\ fraction \\(\%)}	& \makecell{K-peak \\(ns)}	&  \makecell{L-peak \\(ns)} \\
    \hline
    Inner-1	& YCT92	& 630	& 8.8	& 17.6	& 13.0\\
    Inner-2	& YCT2	& 640	& 9.5	& 16.6	& 10.1\\
    Inner-3	& YCN43	& 650	& 9.3	& 13.2	& 10.0\\
    Inner-4	& YCT97	& 640	& 9.2	& 17.3	& 11.4\\
    Inner-5	& YCN46	& 650	& 9.5	& 15.2	& 11.3\\
    Inner-6 & YCN42	& 640	& 9.8	& 13.2	& 9.1\\
    Inner-7	& YCT92	& 640	& 9.3	& 17.6	& 13.0\\
    Inner-8	& YCT2	& 645	& 9.5	& 16.6	& 10.1\\
    Inner-9	& YCN43	& 640	& 9.1	& 13.2	& 10.0\\
    Inner-10& YCT97	& 650	& 9.1	& 17.3	& 11.4\\
    \hline
    \end{tabular}
    \label{tab:inner_tn}
\end{table}

\begin{table}[!t]
    \caption{The cut values on $T_N$ of the pulses for $^{71}$Ge K and L peak for each counting in the outer target. Events with $T_N$ values greater than each cut value are selected out.}
    \centering
    \footnotesize
    \begin{tabular}{|c|c|c|c|c|c|}
    \hline
    \multicolumn{4}{|c|}{Counter Filling} &  \multicolumn{2}{c|}{$T_N$}\\
    \hline
    \makecell{Extraction \\name}	& \makecell{Counter \\name}	& \makecell{Pressure \\(mmHg)}	& \makecell{GeH$_4$ \\ fraction \\(\%)}	& \makecell{K-peak \\(ns)}	&  \makecell{L-peak \\(ns)} \\
    \hline
    Outer-1	& YCN113& 	635	& 9.5	& 13.6	& 9.1 \\
    Outer-2	& YCT3	& 635	& 9.5	& 16.4	& 10.3\\
    Outer-3	& YCNA9	& 640	& 10.5	& 18.8	& 13.2\\
    Outer-4	& YCT9	& 635	& 9.6	& 14.9	& 9.1\\
    Outer-5	& YCN41	& 635	& 10.0	& 13.4	& 10.3\\
    Outer-6	& YCT4	& 630	& 9.0	& 13.2	& 10.2\\
    Outer-7	& YCN113& 	630	& 10.3	& 13.6	& 9.1\\
    Outer-8	& YCT3	& 640	& 9.5	& 16.4	& 10.3\\
    Outer-9	& YCNA9	& 635	& 9.9	& 18.8	& 13.2\\
    Outer-10& YCT9	& 645	& 9.5	& 14.9	& 9.1\\
    \hline
    \end{tabular}
    \label{tab:outer_tn}
\end{table}

We determine the proper energy scaling for each extraction from $^{55}$Fe calibrations. 
The extrapolations were verified with $^{71}$Ge filled counters operated after the BEST extraction counting ended. 
This additional study enabled the comparison between the $^{71}$Ge K and L peak 2~FWHM regions predicted by extrapolation and the true peak positions. 
All counters used in the experiment underwent the performance study.
Details of the counter calibration using $^{71}$Ge isotope are described in Appx.~\ref{app:ge71}.

\subsection{Event selection}\label{sec:event_selection}

Events are further selected to reduce the background that mimics $^{71}$Ge-induced events. 
The event selection algorithm uses the SAGE analysis~\cite{sage1999source}, which has been continuously updated throughout the last 20 years. Some of the updates were reported in~\cite{abdurashitov2009measurement}.

We implement five event selection cuts to remove non-physical signals: 1) flatness cut, 2) shield-open cut, 3) NaI veto cut,  4) setting the energy windows for the Ge L and K peaks and 5) rise-time cut. These cuts effectively remove the background events. 
The first three types of cuts affects only the background level of the counters, and do not affect the selection of real events from $^{71}$Ge decays. 
Therefore the efficiency of such cuts can be considered equal to 100\%. 
The use of such cuts off reduces the background of the counters and, accordingly, the magnitude of the statistical error.

\subsubsection{Flatness cut}

\begin{figure}[thp]
    \centering
    \includegraphics[width=\columnwidth]{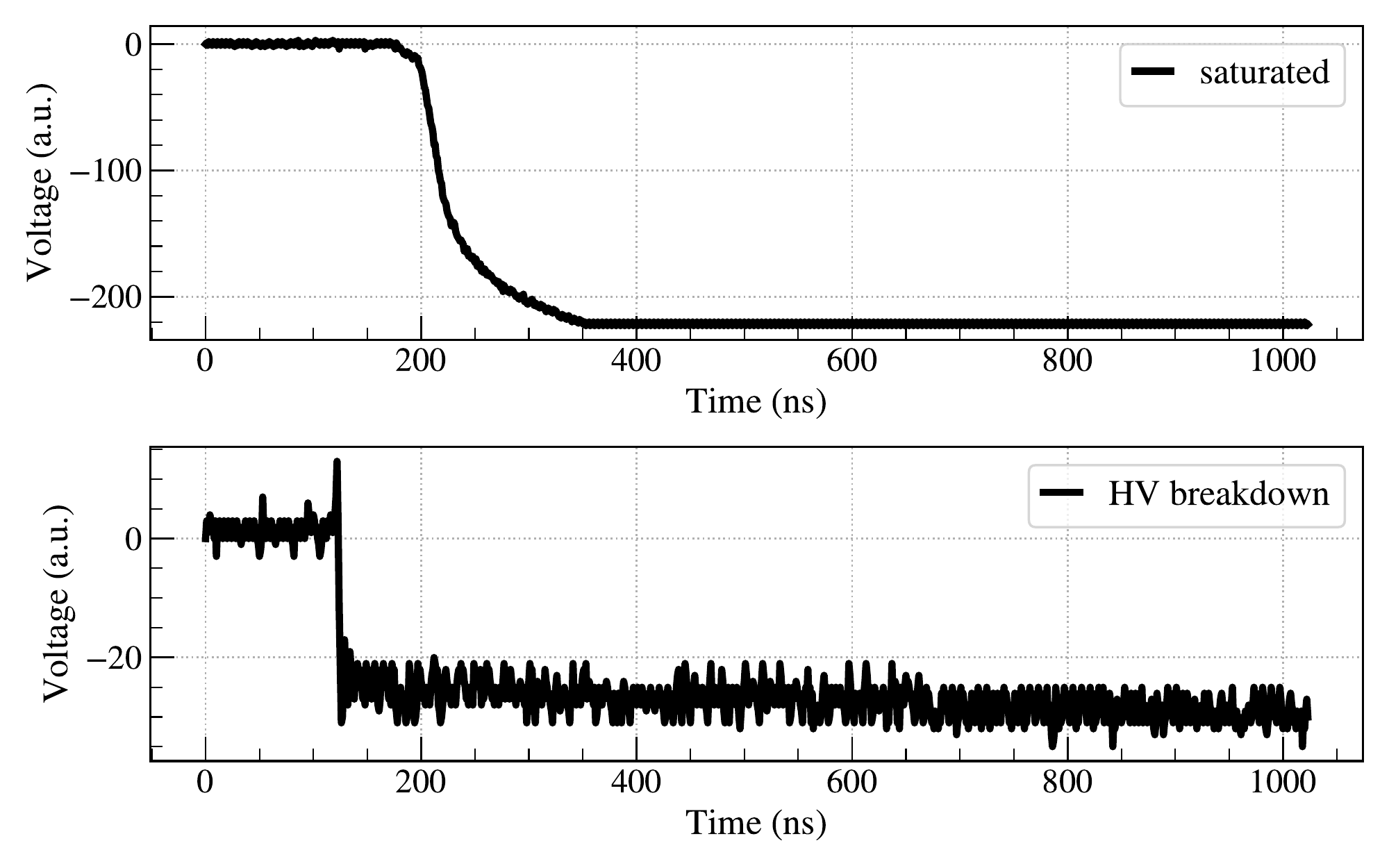}
    \caption{Flat-top pulses measured in proportional counters. \textbf{Top:} Saturated event induced by high-energy $\alpha$-particles. \textbf{Bottom:} Background candidate event that originates from high-voltage breakdown.}
    \label{fig:signal_vs_background_alpha}
\end{figure}

\begin{figure}[thp]
    \centering
    \includegraphics[width=\columnwidth]{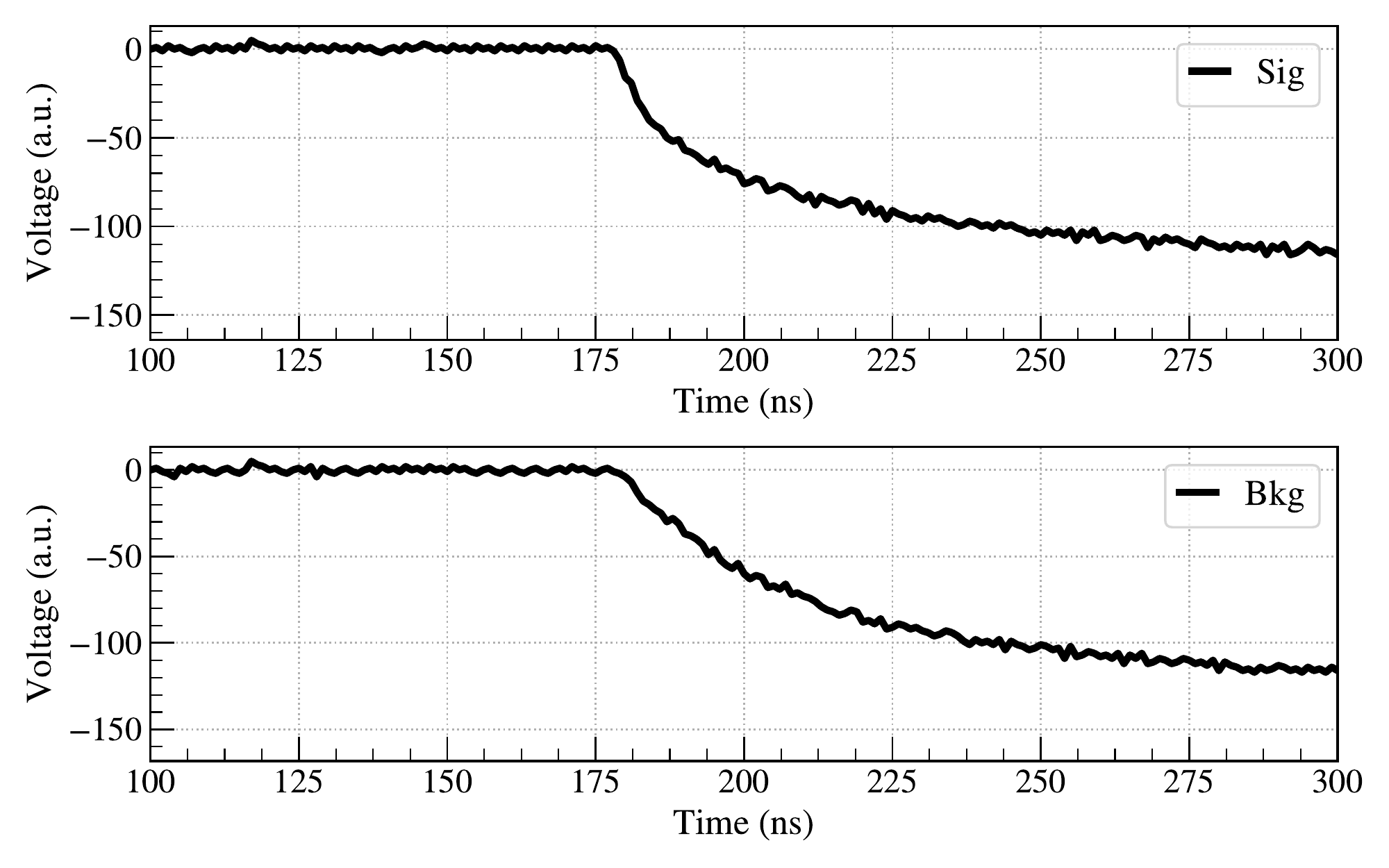}
    \caption{Typical pulse shapes measured in proportional counters. \textbf{Top:} $^{71}$Ge K peak candidate event with $T_N=$ 7.44~ns. \textbf{Bottom:} Background candidate event induced by either Compton scattering or high-energy $\beta$-particles in K peak energy region. The slow background event with extended ionization has much slower fall time ($T_N=$ 30.67~ns) when the pulse begins at $\approx$200~ns than  the true $^{71}$Ge candidate event.}
    \label{fig:signal_vs_background}
\end{figure}

The first step of event selection is to identify two types of background events: saturation waveforms that are mostly from high energy $\alpha$-particles and those that originate from high-voltage breakdown.
The $\alpha$-particles, either from the decay of internal $^{222}$Rn or the natural radioactivity in the counter material, can easily be identified by examining the end of the waveforms. 
Saturated pulses have amplitudes greater than 16~keV and relatively flat shape at the end of the pulse. 
The events from high-voltage breakdown have a characteristic waveform that rises steeply and stays flat at the end. 
While the positive ions from the real $^{71}$Ge events are collected smoothly over time with resulting extended rise time, events from high-voltage breakdown have a characteristic waveform that rises steeply and stays flat at the end after the rising edge.
Because both types of background events have a characteristic shape at the end of the waveform, we analyze the pulses between 500 and 1000~ns after the digitization begins. 
Typical waveforms of an $\alpha$-induced saturated pulse and a background that originates from high-voltage breakdown are shown in Fig.~\ref{fig:signal_vs_background_alpha}.
They are easily distinguishable from the typical pulse shape of true $^{71}$Ge candidate event shown in Fig.~\ref{fig:signal_vs_background}.

\subsubsection{Shield-open cut}

Next, we apply a cut to remove data during periods of expected high backgrounds.
When counters were calibrated, they were exposed to the laboratory atmosphere with an average Rn content of 2~pCi/liter. 
Events produced by $^{222}$Rn daughters can produce false $^{71}$Ge signals that mimic our signal, and hence the effect of external Rn is minimized by making a time cut on the data for 2.6~h after any shield opening.
The estimated effectiveness of this time cut is nearly 100\%.
Any rejected period are considered in our exposure calculation, and there is no efficiency associated with it. 
The uncertainty is found to be negligible~\cite{gavrin2003background}.

\subsubsection{Time cut of Radon chains}

We apply an additional time cut to minimize the number of false $^{71}$Ge signals from internal radon and its daughters~\cite{sage1999solar}. 
The first two decays in the $^{222}$Rn chain are $\alpha$ decays producing off-scale pulses that saturate the detector.
Since $^{222}$Rn decays to $^{210}$Pb~($T_{1/2}-22$-y) typically in about an hour, removing a few hours around each saturated event effectively rejects the false $^{71}$Ge events from the radon chain.
We remove all data from 15~minutes before each saturation event to 3~hours after it~\cite{sage1999solar,abdurashitov2009measurement}. 
The probability that the decays of the elements of this chain will be registered after such cut off is 1.1\% for $^{214}$Pb and 3.5\% for $^{214}$Bi and $^{214}$Po~\cite{gavrin2002effect}
We treat them as the systematic uncertainties related to internal radon.

\subsubsection{NaI veto}

Background events from $\beta$ decays from the counter walls, which are accompanied by $\gamma$ rays are tagged and vetoed by the surrounding NaI detector. 
Since $^{71}$Ge has no $\gamma$ rays associated with its decay other than internal bremsstrahlung, all events coincident with a NaI detector response are eliminated.

\subsubsection{Setting the energy windows for the Ge L and K peaks}

The measure of energy is the integral of the pulse waveform for 800~ns after the pulse onset. The locations of the energy acceptance windows for $^{71}$Ge events in the L and K peaks were set from the $^{55}$Fe calibration as described in Sec.~\ref{sec:calibration}. If the peak position changes from one calibration to the next, then the energy window for event selection is slid linearly in time between the two calibrations. The resolution at each peak is held constant and is set to be the average of the resolutions of $^{55}$Fe for all counter calibrations, scaled to the L- or K-peak energy as described in Sec.~\ref{sec:calibration}. Events are then accepted as candidates only if their energy is within $\pm$ 1~FWHM ((98\% acceptance) of the central peak energy.

\subsubsection{Rise-time cut}

The electron capture process produces point-like ionization in the counter. 
Pulses from the $^{71}$Ge electron capture decay should therefore be fast, and $T_N$ values should be near zero. 
On the other hand, backgrounds from Compton scattering or from high-energy $\beta$-particles traversing the counter produce extended ionizations, leading to slow pulses with large $T_N$. 
Figure~\ref{fig:signal_vs_background} illustrates the difference in pulse shapes of the point-like $^{71}$Ge candidate event with $T_N=7.44$~ns and the background showing the characteristic of extended ionization with $T_N=30.67$~ns. 
Both events fall into the K peak energy region with 9.64~keV, but the background candidate event has much slower fall time when the pulse begins at $\approx$200 ns than the true $^{71}$Ge candidate. 

\begin{figure}[t]
    \centering
    \includegraphics[width=0.8\columnwidth]{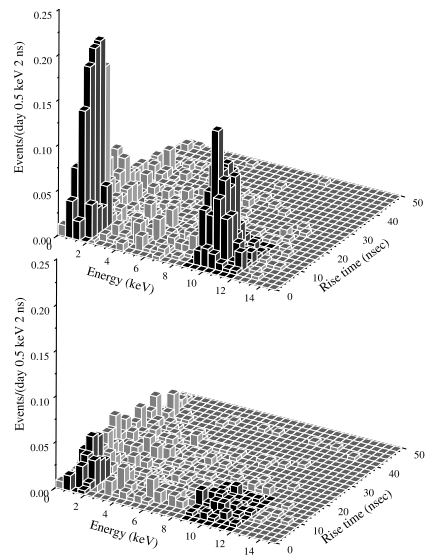}
    \caption{
    \textbf{Upper panel:} energy vs rise-time histogram of all events of the inner target after the shield-open cut observed in all ten exposures during the first 30 days after extraction. The live time is 245~days, and 1364~events are shown. \textbf{Lower panel:} the same histogram for the 481~events that occurred during an equal live-time interval beginning at 40~days after extraction. 
    }
    \label{fig:lego_sph}
\end{figure}

\begin{figure}[t]
    \centering
    \includegraphics[width=0.8\columnwidth]{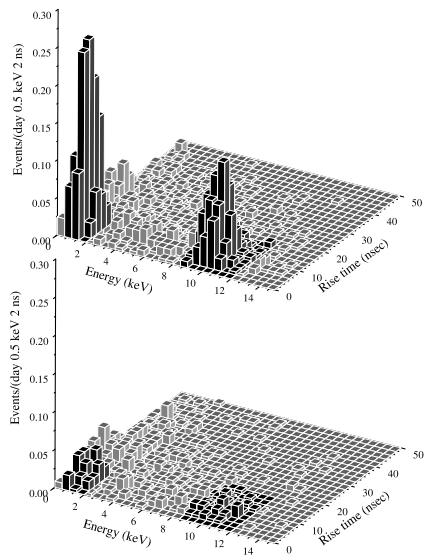}
    \caption{
    \textbf{Upper panel:} energy vs rise-time histogram of all events of the outer target after the shield-open cut observed in all ten exposures during the first 30 days after extraction. The live time is 249~days, and 1387~events are shown. \textbf{Lower panel:} the same histogram for the 504~events that occurred during an equal live-time interval beginning at 40~days after extraction. 
    }
    \label{fig:lego_cyc}
\end{figure}

The cut value for the $T_N$ for each counter is obtained from separate measurements by filling counters with a typical gas mixture (see Tables~\ref{tab:inner_tn} and~\ref{tab:outer_tn}) and adding a trace of active $^{71}$GeH$_4$.  
The counters  were  filled with a mixture  (\%GeH$_4$ $\sim$10\% and pressure $\approx$650~mmHg) very similar to the mixture used in the BEST extractions.
Each proportional counter was installed into the same channel of the counting system with approximately the same high voltage as used in the BEST measurements.
The cut values for the K and L peaks are set such that 4\% of the slowest events in $^{71}$Ge calibration data are excluded.
The derived typical cut values for K and L peaks were 13.2~ns and 9.1~ns respectively. 
However, the cut values vary by detector, and individual $T_N$ values were used for each counter.
The cut values on $T_N$ windows for K and L peak for each counter are given in Tables~\ref{tab:inner_tn} and~\ref{tab:outer_tn}.

\begin {table}[tp]
\caption {A summary of event selection cuts and their individual acceptances.} 
\label{tab:CutSummary} 
\begin{center} 
 \begin{tabular}{ l l  } 
 \hline\hline
 Cut 				& Efficiency (\%)					\\ [0.5ex]  
 \hline
Energy window	&98$\pm1.3$			\\
Rise-time		&96		\\
\hline\hline
\end{tabular}
\end{center}
\end{table}

Table~\ref{tab:CutSummary} summarizes the event selection cuts applied in the BEST analysis and their individual acceptances. 
Figures~\ref{fig:lego_sph} and~\ref{fig:lego_cyc} show the histograms of all events from the inner and the outer zones, after the first four cuts,  for the earlier and the later counting times. 
The number of events outside the peaks is about the same in both panels as these are mainly due to background.
The expected location of the $^{71}$Ge K and L peaks as predicted by the $^{55}$Fe calibrations are shown as darkened. 

\begin{figure}
    \centering
    \includegraphics[width=\columnwidth]{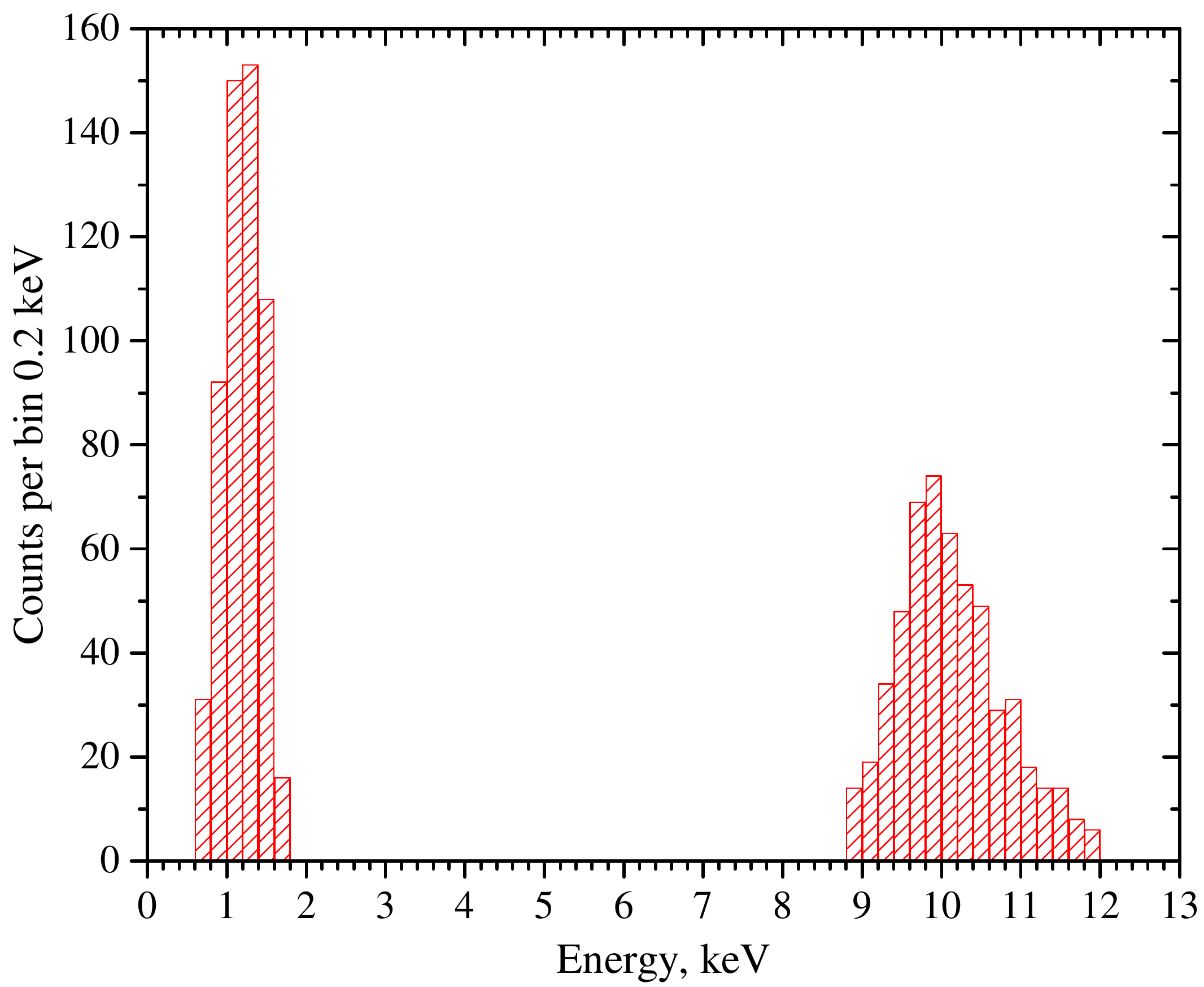}
    \includegraphics[width=\columnwidth]{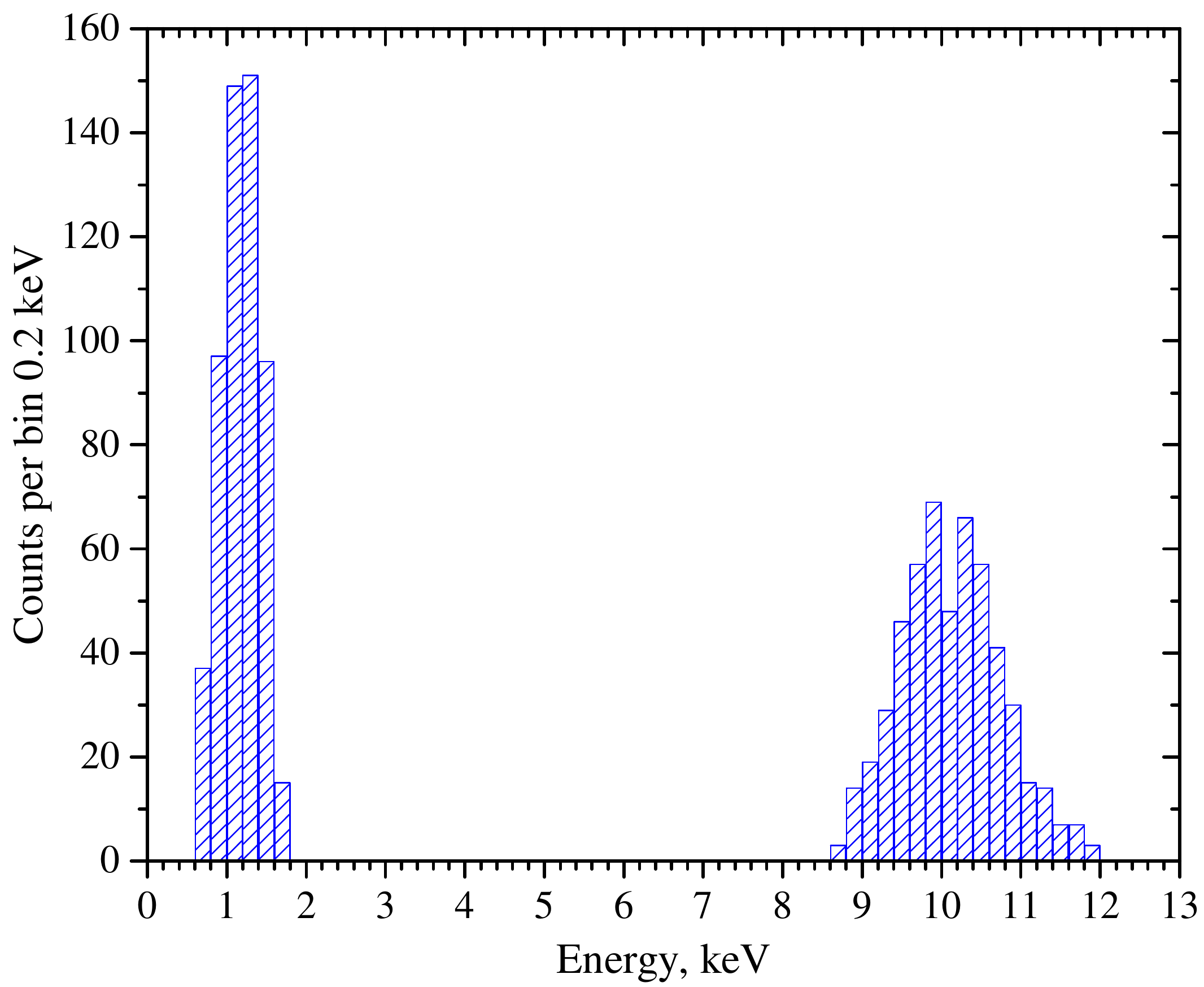}
    \caption{Plots with energy spectra of selected events for whole dataset showing K and L peaks. \textbf{Top:} Energy spectrum of the inner target. \textbf{Bottom:} Energy spectrum of the outer target.}
    \label{fig:e_spectra}
\end{figure}

Figure~\ref{fig:e_spectra} shows the energy spectra of the selected K and L peak candidate events for the inner and the outer targets. 
We note that the event selection windows were determined from post-experiment calibrations, and hence the cut boundary parameters were effectively blind during the analysis.
No additional blindness scheme was implemented.

\subsection{Neutrino-Nucleon Cross Section}

The neutrino-nucleon cross section ($\sigma$) has to be calculated from nuclear physics input. 
The bulk of the cross section is in the ground-state to ground-state transition determined from the matrix element for the decay of $^{71}$Ge.
When the original gallium anomaly was observed, there was concern that the transition strengths to excited states were not fully understood. The cross section was first estimated in the seminal work of Bahcall~\cite{Bahcall1997}. He derived the ground state contribution from the $^{71}$Ge half-life, but the excited state contributions were estimated from charge exchange (i.e. (p,n)) reactions. Bahcall considered a number of possible uncertainties, the largest coming from the excited state contributions due to the quality of the (p,n) data available at that time. For the central value, Bahcall used the best estimate of the transition strength values to the first two excited states and a Q-value of (232.69$\pm$0.15)~keV. Other excited states are too high in energy to contribute for $^{51}$Cr $\nu$'s energy. Bahcall estimated the uncertainty to be the change in $\sigma$ (-1.6/+2.8\%) if one ignores the excited states. Other uncertainties came from forbidden corrections to the beta decay matrix elements (2.3\%) with small uncertainties associated with the $Q_{val}$ (0.05\%), the reaction threshold (0.2\%), and the $^{71}$Ge half life (0.3\%). He considered part of the excited state uncertainty as one-sided, hence the asymmetric uncertainties.  The charge exchange data has been improved~\cite{ejiri1998spin} by recent work~\cite{Frekers2011,Frekers2013,Frekers2015}  indicating that they are not the cause of the discrepancy. However, the excited-state contribution uncertainty is critical because the (p,n) measurements have a significant cancellation between the Gamow-Teller and tensor matrix elements resulting in an underestimate of the transition strengths~\cite{HAXTON1998110}.  The (p,n) cross section could be entirely due to the tensor interaction, in which case (either) excited state contribution to neutrino absorption might be zero.  Unfortunately, there is no direct measurement of the cross section at this energy that is not subject to the caveats of neutrino oscillation physics. 

\begin {table}[tp]
\caption {A summary of the published neutrino-nucleon cross section estimates for $^{71}$Ga in units of $10^{-45}$cm$^2$.} 
\label{tab:CossSectionEstimates} 
\begin{center} 
 \begin{tabular}{ l l l } 
 \hline\hline
 Reference 				& Cross Section			& 	Q-value (keV)					\\ [0.5ex]  
 \hline
\cite{Bahcall1997}		&$5.81^{+0.21}_{-0.16}$		&	232.69(15)				\\
\cite{Frekers2015}		&5.93$\pm$0.14  			&	232.69(15)				\\	
\cite{Barinov2018}		&$5.910\pm0.114$			&	233.5(1.2)					\\
\cite{Kostensalo2019}	& 5.67$\pm$0.06			&	232.49(22)						\\
 \cite{Semenov2020}		& 5.938$\pm$0.116			&	232.443(93)				\\
\hline\hline
\end{tabular}
\end{center}
\end{table}

Kostensalo \textit{et al.}~\cite{Kostensalo2019} used a more recent $Q_{val}=(232.49\pm0.22)$~keV~\cite{ABUSALEEM2011133} and a nuclear shell model calculation with wave functions obtained using recent two-nucleon interactions to calculate the transition strengths. The shell model calculations thus avoid the drawback of using the (p,n) measurements but need to be experimentally confirmed by indirect means, such as state energies and electromagnetic properties. The paper of Semenov \textit{et al.}~\cite{Semenov2020} reproduces Bahcall's approach but uses modern values for the transition strengths and a Q-value of (232.443$\pm$0.093)~keV~\cite{ALANSSARI2016}. This small difference in Q-value between the estimates generates a small uncertainty compared to the other uncertainties in this work and we ignore it. 

The Semenov \textit{et al.} and Kostensalo \textit{et al.} results differ by about 4\%, which  is about 2-3 times larger than the uncertainty estimated for each. Interestingly, the original Bahcall number is half way between these two results with a $\pm$4\% uncertainty. We therefore use the Bahcall $\sigma$ value and the associated conservative uncertainties from his estimate of $[(5.81^{+0.21}_{-0.16})\times10^{-45}$]~cm$^2$. The recent cross section estimates are summarized in Table~\ref{tab:CossSectionEstimates}.

\subsection{Energy-weighted Likelihood Fits}

The time sequence of the candidate $^{71}$Ge events are analyzed with a maximum likelihood method to separate the $^{71}$Ge decay with 11.4~day half-life from remaining background, which is constant over time. 
The likelihood fit takes into account the decay of $^{51}$Cr during the exposure period.

In general, the likelihood function ($\mathcal{L}$) used in the BEST analysis can be written for a single run as

\begin{equation} \label{eq:likelihood}
    \mathcal{L} = e^{-\int_\mathrm{counting~time} P(t,E)dt} \prod_i^l P(t_i,E_i)dt_i~,
\end{equation}

\noindent
where $l$ is the total number of observed events, $t_i$ is the time of event $i$, and $P(t,E)$ is the total probability that an event with energy $E$ will occur at time $t$. If the events that pass all selection criteria are classified as either the signal or the background, the function $P(t,E)$ can be generalized to the form

\begin{equation} \label{eq:function_p}
    P(t,E) = w_p(E)p\epsilon e^{-\lambda t} + w_b (E)b~,
\end{equation}

\noindent
where $p$ and $b$ are the production rate and the background rate, respectively; $\epsilon$ is the overall efficiency; and the terms $w_p(E)$ and $w_b(E)$, called the weight factors, are the energy-dependent probabilities that the event is due to signal or to background, respectively. 
The functional form of $w_p(E)$ must be obtained from an experiment in which only signal is present, and the form of $w_b(E)$ is measured when only background is present. 
It is assumed that both weight factors are independent of time and are normalized such that

\begin{equation} \label{eq:weight_factor_normalization}
    \int_{E_{lo}}^{E_{hi}} w_p(E)dE = \int_{E_{lo}}^{E_{hi}} w_b(E)dE = 1~,
\end{equation}

\noindent
where the integrals extend over the range $E_{lo}$ to $E_{hi}$ used in event selection ($\pm$1 FWHM for this analysis). 

Combining Eq.~\ref{eq:likelihood},~\ref{eq:function_p} and~\ref{eq:weight_factor_normalization}, the likelihood function is generalized to

\begin{equation}\label{eq:modified_likelihood}
\footnotesize
    \mathcal{L} = e^{-p \epsilon \Delta / \lambda-b\tau} \prod_i^l \big[ w_p(E_i)p\epsilon e^{-\lambda t_i} + w_b(E_i)b\big] dt_i dE_i~,
\end{equation}

\noindent
where $\Delta$ is the probability an event will occur while counting is in progress and $\tau$ is the total counting time. 
For maximization purposes, the likelihood function can be arbitrary up to a constant. 
Hence, Eq.~\ref{eq:modified_likelihood} can be divided by $\prod_i^l w_b(E_i)dt_idE_i$ and written as

\begin{equation}
    \mathcal{L} = e^{-p\epsilon\Delta /\lambda -b \tau} \prod_i^l \big[ \frac{w_p(E_i)}{w_b(E_i)}p \epsilon e^{-\lambda t_i} + b\big]~,
\end{equation}

\noindent
which is the standard function that is used by all radiochemical solar neutrino experiments to analyze their counting data where the usual term $e^{-\lambda t_i}$ has been replaced by

\begin{equation}
    \frac{w_p(E_i)}{w_b(E_i)} e^{-\lambda t_i}~.
\end{equation}

The signal and the background weight factors were determined by examining the energy spectra for each counter. 
The use of the energy weight factors shifts the results by values that are quite compatible with the statistical uncertainty.

\subsection{Predicted $^{71}$Ge Production Rates} \label{sec:predicted_rates}

\begin{table*}[th]
    \caption{Values and uncertainties of the terms that enter the calculation of the predicted production rate. All uncertainties are symmetric except for the cross section.}
    \centering
    \begin{tabular}{|l|c|c|c|}
    \hline
    \hline
    & & \multicolumn{2}{c|}{Uncertainty}\\
    \cline{3-4}
    & Value	& Magnitude	& \% \\
    \hline
    Ga density $\rho$ (g Ga/cm$^3$)                     & 6.095	    & 0.002	& 0.033 \\
    Avogadro’s number $N_0$ ($10^{23}$ atoms Ga/mol) 	& 6.0221	& 0.0	& 0.0 \\
    Ga molecular weight $M$ (g Ga/mol)              	& 69.72307	& 0.00013	& 0.0002 \\    
    $^{71}$Ga Atomic abundance $f_1$  (atoms $^{71}$Ga/100 atoms Ga)~\cite{machlan1986absolute}  	& 39.8921 	& 0.0062	& 0.016 \\
    Atomic density $D=\rho N_0 f_1 / M$ ($10^{22}$ atoms $^{71}$Ga/cm$^3$)  	& 2.1001	& 0.0008	& 0.037 \\
    Source activity at reference time  $A$, MCi	        & 3.414	    & 0.008	& 0.23 \\
    Cross section $\sigma$ ($10^{-45}$ cm$^2$/ ($^{71}$Ga atom $^{51}$Cr decay)), Bahcall & 5.81	& +0.21,-0.16	& +3.6,-2.8 \\
    Path length in Ga $<L_{in}>$ (cm)                  & 52.03	    &  0.18	& 0.3 \\
    Path length in Ga $<L_{out}>$ (cm)                  & 54.41	    &  0.18	& 0.3 \\
    \hline 
    \hline
    Predicted production rate ($^{71}$Ge atoms/d), r$_\textrm{In}$	& 69.4	& +2.5,-2.0	& +3.6,-2.8 \\
    Predicted production rate ($^{71}$Ge atoms/d), r$_\textrm{Out}$	& 72.6	& +2.6,-2.1	& +3.6,-2.8 \\
    \hline
    \end{tabular}
    \label{tab:predict_calculation}
\end{table*}

\begin{figure}[t]
    \centering
    \includegraphics[width=1.0\columnwidth]{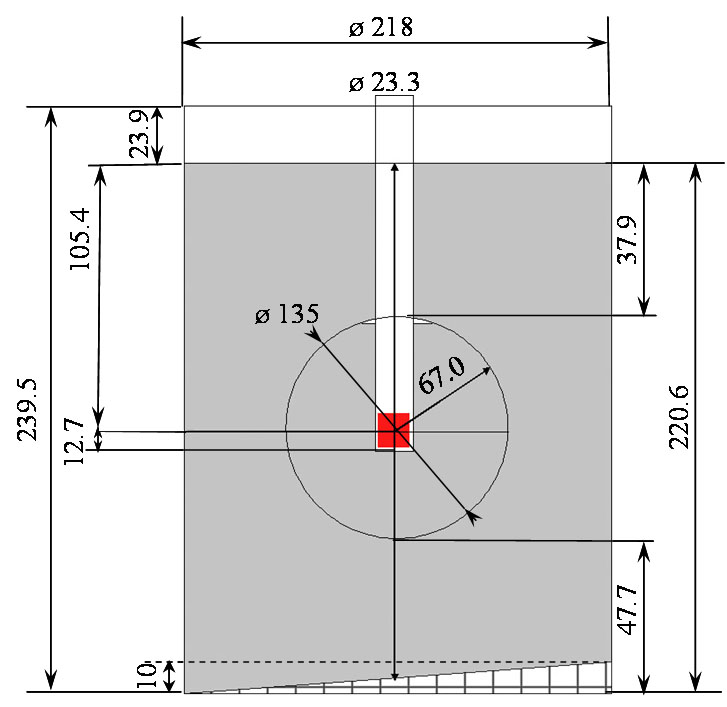}
    \caption{The geometrical dimensions of the Ga targets. All dimensions are in centimeters. There is an empty segment in the spherical vessel due to the thickened wall and the bottom of the re-entrant tube.}
    \label{fig:TargetGeometry}
\end{figure}

The expected $^{71}$Ge production rates depend on various physical constants presented in Table~\ref{tab:predict_calculation} and the geometries of the targets, illustrated in Fig.~\ref{fig:TargetGeometry}.
To account for the non-trivial shapes of the Ga targets, a Monte Carlo integration technique was adapted for the neutrino capture rate calculations in the target zones. 
Calculations with the nominal geometry described in Fig.~\ref{fig:TargetGeometry} yield 1.22545~m$^3$ and 6.5561~m$^3$, which correspond to 7.4691~t and 39.9593~t of liquid Ga. 
The measured masses of the targets are 7.4060~t and 39.9617~t for the inner and outer zones, respectively, with negligible uncertainty for both masses. 

We attribute the discrepancy of the measured and the calculated mass values to the uncertainty in the dimensional measurements, especially in the shell thickness of the spherical volume. 
In a study to explore how sensitive the Monte Carlo results are to various alterations in geometry, modifying the spherical shell thickness from the nominal value of 0.50~cm to 0.68~cm was found to neutralize the difference in the mass values.
We include the error of inaccurate knowledge of the sphere shell thickness as a systematic uncertainty of the path length, which for the sphere amounts to 0.4\% and for the cylinder is negligible. 
Accounting also for the uncertainties in gallium weighing, we use (7.4691$\pm$0.0631)~t and (39.9593$\pm$0.0024)~t as the nominal masses of the inner and the outer targets. 

The effective path length in a volume, or the average path length the $\nu_e$ takes through the Ga target, is calculated by the integral

\begin{equation}
    <L> = \int_V \frac{1}{4\pi d^2}~d\vec{x}~,
\end{equation}

\noindent
where $d$ is the distance between emission in the source and absorption in the Ga targets

We carry out the Monte Carlo integration to yield the path length values for the inner and the outer volumes. Taking into account both the statistical uncertainty due to the $10^7$ sample Monte Carlo integration and the systematic uncertainty from the dimensional uncertainties of the apparatus, the average path lengths for the two volumes are

\begin{equation}
    \begin{split}
        <L>_{in}&=(52.03\pm0.18)~\mathrm{cm~, and} \\
        <L>_{out}&=(54.41\pm0.18)~\mathrm{cm}~.
    \end{split}
\end{equation}

The neutrino capture rate $r_j$ in the volume $V_j$ can be written as

\begin{equation} \label{eqn:neutrino_capture_rate}
\begin{split}
r_j &= \sum_k^4 \int_{V_j} F P_{ee}(\vec{x},E_{\nu_k}) \sigma_k n d\vec{x} \\
    &\approx V_0 \frac{1}{N} \sum_k^4 \sum_{i=1}^{N} \frac{p_k P_{ee}(d,E_{\nu_k})}{d^2}  \Theta_j(\vec{x}_i)~,
\end{split}
\end{equation}

\noindent 
where $F$ is the flux of $\nu_e$, $P_{ee}(\vec{x},E_{\nu_k})$ is the oscillation survival probability of $\nu_e$ with energy $E_{\nu_k}$, $p_k$ is the branching ratio for the $\nu_e$ with energy $E_{\nu_k}$, $\sigma_k$ is the cross section, $n$ is the $^{71}$Ga number density, and $\Theta(\vec{x}_i)=1$ if $\vec{x}_i$ is within $V$ and 0 otherwise. 
The summation over $k$ is for the four neutrino energies and their branching ratios 747~keV~(81.63\%), 427~keV~(8.95\%), 752~keV~(8.49\%) and 432~keV~(0.93\%)~\cite{acero2008limits}.
The summation over $i$ is for the Monte Carlo integration over $N$ samples. 
For the no oscillation case, the survival probability at a short distance is simply $P_{ee}(d)=1$. 
In the case where only one sterile mass eigenstate $m_4$ is introduced to the standard three active mass states and $m_4 \gg m_1, m_2, m_3$, the survival probability of any active neutrino at a short distance $d$ is governed by the two-neutrino oscillation model

\begin{equation}\label{eqn:survival_probability}
P_{ee}(E_\nu,d) = 1 - \mbox{sin}^22\theta \mbox{sin}^2 \left(1.27 \frac{\Delta m^2[\mbox{eV}^2] d[\mbox{m}]}{E_{\nu}[\mbox{MeV}]} \right)~,
\end{equation}

\noindent
where $\Delta m^2$ is the mass squared difference between the active and the inactive state, $\theta$ is the mixing angle, and $E_\nu$ is the neutrino energy.

Using the source activity of (3.414$\pm$0.008)~MCi, the neutrino-nucleus cross section, and combining the uncertainty in quadrature, Eq.~\ref{eqn:neutrino_capture_rate} gives the predicted production rates of

\begin{equation}
\begin{split}
    r&_\textrm{In-predic} = (69.4^{+2.5}_{-2.0})~\textrm{atom/d}~, \\
    r&_\textrm{Out-predic} = (72.6^{+2.6}_{-2.1})~\textrm{atom/d}~.
\end{split}
\end{equation}

The uncertainties are dominated by the cross-section uncertainty, and thus are strongly correlated in these predictions for the two volumes.

\subsection{Measured $^{71}$Ge Production Rates}

The results of the likelihood fits for individual runs and the combined fits are presented in Tables~\ref{tab:inner_measured} and~\ref{tab:outer_measured}. 
For each run and also for the combined runs, the likelihood fits to the time distribution of the candidate events were performed as described in Refs.~\cite{sage1999source,sage2006argon}.

The background from solar neutrino capture conforms to (66.1$\pm$3.1)~solar neutrino unit~(SNU \footnote{1 SNU = 1 interaction per 10$^{36}$ target atoms per second})  ~\cite{abdurashitov2009measurement} as explained below, and was taken into account when determining the production rate from the source. 
In general, the contribution from the solar neutrino capture is 0.6 and 3.2 counts for the inner and the outer targets. 
These are 0.8\% and 4.6\% of the total numbers fit to $^{71}$Ge, respectively.
Due to the inefficiency of extractions, there are also residual $^{71}$Ge atoms that carryover from one extraction to the next. 
The contribution from the carryover atom was typically less than 1 count for each volume and was also taken into account.

\begin{table*}[htp]
    \caption{A summary of the likelihood fits for the production rate for the inner target from each extraction and the combined fit of all extractions, in K and L peak independently and combined. The production rates for each exposure for the K and L-peaks are referred to its starting time, taking into account the total detection efficiency. The production rate for their combined result is referred to the reference time. All production rates for the K+L-peak are referred to the reference time. The stated uncertainties are statistical and given in 68\% confidence. The Cram\'{e}r-von Mises statistics $Nw^2$ measures the goodness of fit of the sequence of event times. The probability was inferred from $Nw^2$ by simulation.}
    \centering
    \def\arraystretch{1.3}
    \begin{tabular}{|c|c|c||c|c|c||c||c|c|}
    \multicolumn{1}{l}{K-peak} \\
    \hline
    Extraction	& \makecell{Number of \\ candidate \\ events}	& \makecell{Number \\ fit to \\ $^{71}$Ge}	& \makecell{$^{51}$Cr source \\ production}	& \makecell{Solar $\nu$ \\production}	 & Carryover	& \makecell{$^{71}$Ge Production \\ decay rate  \\ (atoms/day)} & $Nw^2$ & \makecell{Probability \\(\%)}	\\
    \hline
    Inner-1	& 88	& 88.0	& 87.6	& 0.4	& 0.0	& 48.2$^{+6.3}_{-4.2}$ & 0.110	& 49 \\
    Inner-2	& 81	& 70.2	& 68.1	& 0.4	& 1.7	& 40.5$^{+5.6}_{-6.8}$ & 0.081	& 58 \\
    Inner-3	& 69	& 61.2	& 60.1	& 0.3	& 0.8	& 39.0$^{+5.9}_{-6.1}$ & 0.124	& 35 \\
    Inner-4	& 65	& 65.0	& 63.9	& 0.3	& 0.8	& 37.5$^{+5.6}_{-4.2}$ & 0.057	& 82 \\
    Inner-5	& 57	& 32.5	& 31.6	& 0.3	& 0.6	& 19.5$^{+3.6}_{-4.6}$ & 0.107	& 23 \\
    Inner-6	& 44	& 36.4	& 35.7	& 0.2	& 0.5	& 22.0$^{+3.7}_{-4.4}$ & 0.026	& 96 \\
    Inner-7	& 43	& 23.0	& 22.4	& 0.3	& 0.3	& 14.0$^{+3.1}_{-3.9}$ & 0.048	& 70 \\
    Inner-8	& 28	& 15.9	& 15.3	& 0.3	& 0.3	& 8.9$^{+2.6}_{-3.3}$ &  0.160	& 11 \\
    Inner-9	& 40	& 6.3	& 5.8	& 0.3	& 0.2	& 3.4$^{+2.0}_{-2.9}$ &  0.071	& 54 \\
    Inner-10& 28	& 13.3	& 12.9	& 0.3	& 0.1	& 7.8$^{+2.3}_{-3.0}$ &  0.031	& 91 \\
    \hline
    Comb. K	& 543	& 411.6	& 403.2	& 3.0	& 5.4	& 58.5$^{+3.4}_{-3.5}$ & 0.095	& 44 \\
    \hline
    \multicolumn{1}{l}{L-peak} \\
    \hline
    Extraction	& \makecell{Number of \\ candidate \\ events}	& \makecell{Number \\ fit to \\ $^{71}$Ge}	& \makecell{$^{51}$Cr source \\ production}	& \makecell{Solar $\nu$ \\production}	 & Carryover	& \makecell{$^{71}$Ge Production \\ decay rate  \\ (atoms/day)} & $Nw^2$ & \makecell{Probability \\(\%)}	\\
    \hline
    
    Inner-1	& 92	& 89.5	& 89.2	& 0.4	& 0.0	& 51.4$^{+8.3}_{-4.1}$	& 0.353	& 6 \\
    Inner-2	& 48	& 42.5	& 40.8	& 0.4	& 1.4	& 28.0$^{+5.2}_{-5.7}$	& 0.023	& 99 \\
    Inner-3	& 63	& 56.2	& 55.1	& 0.3	& 0.8	& 35.8$^{+6.1}_{-5.8}$	& 0.079	& 58 \\
    Inner-4 & 	28	& 19.5	& 19.0	& 0.2	& 0.3	& 23.3$^{+6.2}_{-7.4}$	& 0.101	& 36 \\
    Inner-5	& 77	& 27.5	& 26.6	& 0.3	& 0.6	& 16.4$^{+3.4}_{-4.6}$	& 0.046	& 74 \\
    Inner-6 & 	37	& 11.6	& 11.2	& 0.2	& 0.2	& 13.7$^{+4.6}_{-6.0}$	& 0.115	& 20 \\
    Inner-7 & 	48	& 22.1	& 21.5	& 0.3	& 0.3	& 14.4$^{+3.5}_{-4.5}$	& 0.142	& 9 \\
    Inner-8	& 31	& 17.6	& 17.0	& 0.3	& 0.3	& 9.4$^{+2.3}_{-3.1}$	& 0.066	& 53 \\
    Inner-9	& 66	& 17.4	& 16.9	& 0.3	& 0.2	& 9.9$^{+2.9}_{-3.7}$	& 0.209 & 	5 \\
    Inner-10& 60	& 11.4	& 11.0	& 0.3	& 0.2	& 6.3$^{+2.4}_{-3.2}$	& 0.085	& 44 \\
    \hline
    Comb. L	& 550	& 311.3	& 304.0	& 3.0	& 4.3	& 50.7$^{+3.7}_{-3.7}$	& 0.07	& 48 \\
    \hline
    \multicolumn{1}{l}{K+L-peak} \\
    \hline
    Extraction	& \makecell{Number of \\ candidate \\ events}	& \makecell{Number \\ fit to \\ $^{71}$Ge}	& \makecell{$^{51}$Cr source \\ production}	& \makecell{Solar $\nu$ \\production}	 & Carryover	& \makecell{$^{71}$Ge Production \\ decay rate  \\ (atoms/day)} & $Nw^2$ & \makecell{Probability \\(\%)}	\\
    \hline
    Inner-1	& 180	& 176.3	& 175.5	& 0.8	& 0.0	& 49.4$^{+4.0}_{-4.2}$& 0.398	& 6 \\
    Inner-2	& 129	& 111.5	& 107.7	& 0.8	& 3.1	& 44.9$^{+5.6}_{-5.9}$& 0.065	& 70 \\
    Inner-3	& 132	& 117.6	& 115.3	& 0.7	& 1.6	& 62.9$^{+7.1}_{-7.4}$& 0.056	& 76 \\
    Inner-4	& 93	& 87.3	& 85.6	& 0.6	& 1.1	& 73.3$^{+8.0}_{-8.6}$& 0.077	& 59 \\
    Inner-5	& 134	& 60.2	& 58.4	& 0.6	& 1.2	& 49.8$^{+7.7}_{-8.2}$& 0.033	& 92 \\
    Inner-6	& 81	& 48.8	& 47.7	& 0.4	& 0.7	& 69.5$^{+11.0}_{-12.0}$& 0.067	& 49 \\
    Inner-7	& 91	& 45.0	& 43.9	& 0.5	& 0.6	& 64.6$^{+11.6}_{-12.6}$& 0.127	& 13 \\
    Inner-8	& 59	& 33.6	& 32.4	& 0.6	& 0.6	& 53.8$^{+11.0}_{-12.2}$& 0.037	& 87 \\
    Inner-9	& 106	& 23.7	& 22.7	& 0.6	& 0.4	& 49.9$^{+14.9}_{-16.5}$& 0.164	& 10 \\
    Inner-10& 88	& 25.2	& 24.3	& 0.6	& 0.3	& 69.1$^{+17.3}_{-19.4}$& 0.108	& 26 \\
    \hline
    Comb. K+L& 	1093& 724.0	 & 708.2& 6.1	&  9.7	& 54.9$^{+2.4}_{-2.5}$& 0.099	& 28 \\
    \hline
    \end{tabular}
    \label{tab:inner_measured}
\end{table*}

\begin{table*}[htp]
    \caption{A summary of the likelihood fits for the production rate for the outer target from each extraction and the combined fit of all extractions, in K and L peak independently and combined. The production rates for each exposure for the K and L-peaks are referred to its starting time, taking into account the total detection efficiency. The production rate for their combined result is referred to the reference time. All production rates for the K+L-peak are referred to the reference time. The stated uncertainties are statistical and given in 68\% confidence. The  Cram\'{e}r-von Mises  statistics $Nw^2$ measures the goodness of fit of the sequence of event times. The probability was inferred from $Nw^2$ by simulation.}
    \centering
    \def\arraystretch{1.3}
    \begin{tabular}{|c|c|c||c|c|c||c||c|c|}
    \multicolumn{1}{l}{K-peak} \\
    \hline
    Extraction	& \makecell{Number of \\ candidate \\ events}	& \makecell{Number \\ fit to \\ $^{71}$Ge}	& \makecell{$^{51}$Cr source \\ production}	& \makecell{Solar $\nu$ \\production}	 & Carryover	& \makecell{$^{71}$Ge Production \\ decay rate  \\ (atoms/day)} & $Nw^2$ & \makecell{Probability \\(\%)}	\\
    \hline
    Outer-1	& 89	& 77.3	& 75.4	& 1.8	& 0.1	& 46.8$^{+6.1}_{-7.5}$ & 0.051	& 83\\
    Outer-2	& 99	& 89.1	& 86.3	& 1.8	& 1.0	& 49.1$^{+5.1}_{-7.5}$& 0.070	& 67\\
    Outer-3	& 62	& 48.1	& 46.1	& 1.3	& 0.7	& 34.3$^{+5.7}_{-8.2}$& 0.158	& 20\\
    Outer-4	& 64	& 55.2	& 53.0	& 1.7	& 0.5	& 31.2$^{+4.6}_{-6.1}$& 0.037	& 90\\
    Outer-5	& 50	& 28.4	& 26.2	& 1.7	& 0.5	& 15.3$^{+2.5}_{-5.0}$& 0.148	& 9\\
    Outer-6	& 50	& 43.6	& 41.6	& 1.6	& 0.4	& 25.4$^{+3.1}_{-5.4}$& 0.085	& 53\\
    Outer-7	& 28	& 20.5	& 18.7	& 1.6	& 0.2	& 11.6$^{+1.7}_{-4.2}$& 0.227	& 6\\
    Outer-8	& 33	& 25.0	& 22.9	& 1.9	& 0.2	& 12.1$^{+1.7}_{-4.0}$& 0.130	& 25\\
    Outer-9	& 23	& 10.6	& 9.1	& 1.4	& 0.1	& 6.6$^{+0.0}_{-4.0}$& 0.055	& 63	\\
    Outer-10	& 27	& 16.7	& 14.9	& 1.7	& 0.1	& 9.0$^{+1.6}_{-4.0}$& 0.043	& 78	\\
    \hline
    Comb. K	& 524	& 415.4	& 395.1	& 16.6	& 3.7	& 59.4$^{+3.6}_{-3.7}$& 0.066	& 58\\
    \hline
    \multicolumn{1}{l}{L-peak} \\
    \hline
    Extraction	& \makecell{Number of \\ candidate \\ events}	& \makecell{Number \\ fit to \\ $^{71}$Ge}	& \makecell{$^{51}$Cr source \\ production}	& \makecell{Solar $\nu$ \\production}	 & Carryover	& \makecell{$^{71}$Ge Production \\ decay rate  \\ (atoms/day)} & $Nw^2$ & \makecell{Probability \\(\%)}	\\
    \hline
    Outer-1	& 92	& 51.5	& 49.6	& 1.8	& 0.1	& 32.1$^{+6.9}_{-9.1}$& 0.216	& 11\\
    Outer-2 & 75	& 75.0	& 72.6	& 1.5	& 0.9	& 48.1$^{+7.5}_{-4.9}$& 0.035	& 95\\
    Outer-3	& 54	& 45.9	& 43.6	& 1.5	& 0.8	& 28.4$^{+4.2}_{-6.2}$& 0.202	& 15\\
    Outer-4 & 34	& 27.0	& 25.9	& 0.8	& 0.3	& 30.4$^{+6.0}_{-8.5}$& 0.159	& 18\\
    Outer-5	& 70	& 35.6	& 33.3	& 1.8	& 0.5	& 18.6$^{+2.6}_{-5.3}$& 0.045	& 74\\
    Outer-6 & 47	& 18.3	& 17.3	& 0.8	& 0.2	& 22.1$^{+5.1}_{-8.1}$& 0.147	& 9\\
    Outer-7 & 41	& 17.2	& 15.5	& 1.5	& 0.2	& 10.2$^{+2.1}_{-4.6}$& 0.071	& 47\\
    Outer-8	& 35	& 18.0	& 15.9	& 1.9	& 0.2	& 8.1$^{+1.4}_{-3.8}$& 0.060	& 58\\
    Outer-9	& 43	& 9.1	& 7.3	& 1.7	& 0.1	& 4.6$^{+1.5}_{-4.0}$& 0.080	& 44\\
    Outer-10& 54	& 14.6	& 12.6	& 1.9	& 0.1	& 7.1$^{+1.6}_{-4.0}$& 0.038	& 85\\
    \hline
    Comb. L		& 545	& 320.6		& 302.0		& 15.3		& 3.3		& 50.9$^{+3.9}_{-3.9}$& 0.062	& 54\\
    \hline
    \multicolumn{1}{l}{K+L-peak} \\
    \hline
    Extraction	& \makecell{Number of \\ candidate \\ events}	& \makecell{Number \\ fit to \\ $^{71}$Ge}	& \makecell{$^{51}$Cr source \\ production}	& \makecell{Solar $\nu$ \\production}	 & Carryover	& \makecell{$^{71}$Ge Production \\ decay rate  \\ (atoms/day)} & $Nw^2$ & \makecell{Probability \\(\%)}	\\
    \hline
    Outer-1	& 181	& 133.4	& 129.6	& 3.7	& 0.1	& 41.1$^{+5.2}_{-5.3}$& 0.191	& 18\\
    Outer-2 & 174	& 163.8	& 158.6	& 3.3	& 1.9	& 63.6$^{+5.5}_{-5.7}$& 0.065	& 73\\
    Outer-3	& 116	& 92.5	& 88.2	& 2.8	& 1.5	& 51.4$^{+6.9}_{-7.3}$& 0.123	& 32\\
    Outer-4	& 98	& 82.3	& 78.9	& 2.5	& 0.8	& 66.6$^{+9.2}_{-9.8}$& 0.045	& 84\\
    Outer-5	& 120	& 64.0	& 59.5	& 3.5	& 1.0	& 46.9$^{+7.2}_{-7.9}$& 0.068	& 48\\
    Outer-6 & 97	& 62.3	& 59.3	& 2.6	& 0.4	& 87.3$^{+12.3}_{-13.2}$& 0.095	& 30\\
    Outer-7 & 69	& 38.0	& 34.4	& 3.2	& 0.4	& 50.4$^{+9.6}_{-10.6}$& 0.132	& 13\\
    Outer-8	& 68	& 43.4	& 39.2	& 3.9	& 0.4	& 59.7$^{+10.8}_{-11.7}$& 0.072	& 50\\
    Outer-9	& 66	& 20.2	& 17.0	& 3.0	& 0.2	& 43.0$^{+13.5}_{-15.3}$& 0.044	& 80\\
    Outer-10& 81	& 31.8	& 28.0	& 3.6	& 0.2	& 78.8$^{+18.1}_{-20.0}$& 0.028	& 96\\
    \hline
    Comb. K+L& 1069	& 738.8	& 699.8	& 32.2	& 6.8	& 55.6$^{+2.6}_{-2.7}$& 0.079	& 32\\
    \hline
    \end{tabular}
    \label{tab:outer_measured}
\end{table*}

The best fit production rates of $^{71}$Ge from the combined data analysis for each target at the reference time are

\begin{equation}
\begin{split}
	r&_\textrm{In-fit} = (54.9\pm 2.5) ~\textrm{atoms/d}~,\\
	r&_\textrm{Out-fit} = (55.6 \pm 2.7) ~\textrm{atoms/d}~.
\end{split}
\end{equation}

\noindent
The stated uncertainty is statistical and is given with 68\% confidence.
The fit values of $^{71}$Ge half-life from the inner and the outer targets are ($11.11\pm 0.69$)~d and ($11.05\pm0.72$)~d. These are in good agreement with the accepted half-life of ($11.43\pm0.03$)~d~\cite{hampel1985}.

\begin{table*}[th]
    \caption{Summary of the contributions to the systematic uncertainty in the measured neutrino capture rate.}
    \centering
    \begin{tabular}{|l|c|}
    \hline
    Origin of uncertainty	& Uncertainty (\%) \\
    \hline
    Chemical Extraction Efficiency & \\
    \hspace{5mm} Efficiency of extraction from Ga metal, $\delta_{G1}$  & $\pm1.0$ \\
    \hspace{5mm} Efficiency of synthesis into GeH$_4$, $\delta_{G2}$    & $\pm1.3$ \\
    \hspace{5mm} Carrier carryover, $\delta_{G3}$    & negligible \\
    \hspace{5mm} Mass of Ga, $\delta_{G4}$    & negligible \\
    \multicolumn{1}{|r|}{Chemical Extraction Subtotal}                  & $\pm 1.6$ \\  
    Counting Efficiency & \\
    \hspace{5mm} Calculated Efficiency                                  & \\
    \hspace{8mm} Volume Efficiency, $\delta_{C1}$                       & -1.5, +1.8 \\
    \hspace{8mm} Peak Efficiency, $\delta_{C2}$                         & $\pm 1.3$ \\
    \hspace{8mm} \makecell[l]{Simulations to adjust for counter filling, \\Monte Carlo interpolation, $\delta_{C3}$} & $\pm 0.6$ \\
    \hspace{5mm} Calibration Statistics                                 & \\
    \hspace{8mm} Gain Variation, $\delta_{C4}$                          & $+ 0.4$ \\
    \hspace{8mm} Resolution, $\delta_{C5}$                              & $\pm 0.3$ \\
    \hspace{8mm} Centroid, $\delta_{C6}$                                & $\pm 0.1$ \\
    \hspace{8mm} Rise time cut, $\delta_{C7}$ & negligible \\
    \multicolumn{1}{|r|}{Counting Efficiency Subtotal}                  & -2.1, +2.3 \\  
    Background Discrimination & \\
    \hspace{5mm} Residual Radon after Time Cuts, $\delta_{N1}$          & $-0.04$ \\
    \hspace{5mm} Solar Neutrino Background, $\delta_{N2}$               & $\pm0.20$ \\
    \hspace{5mm} $^{71}$Ge carryover, $\delta_{N3}$                     & $\pm 0.05$ \\
    \multicolumn{1}{|r|}{Background Discrimination Subtotal}            & $\pm 0.2$ \\ 
    \hline
    \hline
    Total Systematic Uncertainty                                        & -2.7, +2.9 \\
    \hline
    \end{tabular}
    \label{tab:systematic_uncertainty}
\end{table*}

\subsection{Systematic Studies} \label{sec:syst_uncer}
The systematic uncertainties are categorized into four basic types: 1) uncertainties related to the chemical extraction efficiency~($\delta_G$), 2) uncertainties in the counting efficiency~($\delta_C$), 3) uncertainties arising from background phenomena ($\delta_N$), and the average path uength uncertainty. 
 The uncertainties related to the measurement are summarized in Table~\ref{tab:systematic_uncertainty}, while item 4) is related to the prediction.

\subsubsection{Extraction Efficiencies}\label{sec:extraction_efficiency}

\noindent
\textbf{i. Extraction efficiency from Ga metal, $\delta_{G1}$}

Special Ge carriers were produced with a known number of $^{71}$Ge contents~\cite{sage1999solar}.
These carriers were added to the reactors, and the numbers of $^{71}$Ge atoms extracted were counted and compared to the added numbers of $^{71}$Ge atoms. 
The ratio of extracted and added Ge atoms determines the extraction efficiency.

Two types of carrier with different enrichment in Ge isotopes $^{72}$Ge and $^{76}$Ge were used for the experiment. 
Both ligatures were made in 2015 and were used in preliminary solar measurements.
Ge extraction efficiency was determined by a mass spectrometric method of extraction efficiency determination, which is based on isotopic measurements~\cite{CLEVELAND201541}.
In this method, the efficiency of extraction of Ge from the Ga is calculated from the measured mass and isotopic content of added and extracted Ge and  all relevant uncertainties (uncertainty in the total mass of Ge in the carrier, uncertainty in the Ge isotopic fractions of the carrier, uncertainty in the mass of Ge in the extracted sample, and uncertainty in the Ge isotopic fractions of the measured sample) are take into account.
The uncertainty of the extraction efficiency is $\delta_{G1}=\pm 1.0\,\%$. 
\\
\\
\noindent
\textbf{ii. Efficiency of synthesis into GeH$_4$,  $\delta_{G2}$}

An uncertainty in how much  extracted carrier has been synthesized into GeH$_{4}$. 
The amount of germanium entering the counter is calculated from the volume of synthesized germane. 
The volume of germane is determined using the calibrated stem of the Toepler pump, which is the main apparatus in the gas pumping system and the preparation of the working mixture of the counter. 
In the Topler pump, there are two calibrated stems with measured volumes. 
Their volumes sizes were obtained by a large series of calibration measurements with uncertainty not higher than 1\%. 
The calculation is based on the difference in mercury levels in the evacuated part of the Toepler pump, which is a manometric tube with a linear scale, and a calibrated stems of a precisely known volume, in which germane is located before being placed in the counter. 
Division value of the vertical scale of the pressure gauge tube is 1~mm, and the measurement error is half of 1 division, i.e. 0.5 mm. 
Measurement of the volume of germane is carried out in two calibrated volumes of the Topler
pump, that reduces the measurement error to $\delta_{G2} = \pm 1.3$\%.
\\
\\
\noindent
\textbf{iii. Carrier carryover,  $\delta_{G3}$}

The uncertainty of the amount of residual Ge carrier remaining from previous extractions $\delta_{G3}$ with the mass spectrometric analysis, which is the part of the uncertainty in $\delta_{G1}$. 
Therefore, we assign negligible uncertainty to $\delta_{G3}$.
\\
\\
\noindent
\textbf{iv. Mass of Ga,  $\delta_{G4}$}

The gallium target mass affects the results of measurements of the neutrino capture rate indirectly via neutrino path length from a radioactive source through the gallium target in two zones. 
This uncertainty $\delta_{G4}$ is taken into account in determining the average neutrino path length in the gallium targets, which is given in Sec.~\ref{sec:predicted_rates}.

\subsubsection{Counting Efficiencies}

Uncertainties associated with the counting efficiencies of pulses from $^{71}$Ge decays in proportional counters constitute the second group of uncertainties, $\delta_C$.
\\
\\
\noindent
\textbf{i. Volume efficiency,  $\delta_{C1}$}

The volume efficiencies of the counters were measured at the end of the BEST measurements using two isotopes, $^{37}$Ar and $^{71}$Ge. 
Here, the volume efficiency is defined as the probability that the decay of a radioactive atom in the volume of a proportional counter would produce a detectable signal~\cite{sage1999solar}. 
This efficiency was calculated for each detector by comparing the measured count rate in each count to the count rate in the high-efficiency standardization counter. 
A purified $^{37}$Ar gas was mixed with 90\% Ar plus 10\% CH$_4$, and the mixture is placed into counters under test. 
All counters were filled in turn with the same gas mixture. 
Three spectra were taken from each counter with each filling: the first with the voltage adjusted so that the $^{37}$Ar K peak is set at the middle of the scale to see all pulses with energies above the K peak, the second that both L and K peaks were on the scale, and the third with the L peak, whose energy is 270~eV, at about mid-scale. 
After the measurements, the mixture is transferred to the specially designed standardization counter with a volume efficiency of (99.5$\pm$0.6)\%, with a very high transfer efficiency $E_\mathrm{transfer} > 99.5$\%. 
The standardization counter is brought to about the same pressure as in the test counter and the same spectra are measured. 
The ratio between the event rate above a certain threshold in the test counter, $R_\mathrm{test}$, is then compared with the event rate above the same threshold in the standardization counter, $R_\mathrm{test}$. 
After minor corrections, the volume efficiency of the counter is given by $\epsilon_V = 0.995 R_\mathrm{test}E_\mathrm{transfer}D/R_\mathrm{standard}$ where $D$ is the decay factor of the $^{37}$ between the two measurements~\cite{sage1999solar}. 
Additional measurements with $^{71}$GeH$_4$ mixed with Xe-GeH$_4$ were carried out using a similar technique described above for verification.
The uncertainty assigned to volume efficiency is $\delta_{C1} = +1.5 / -1.3$\%.
Considering that four counters were used twice in the BEST measurements this counting systematic uncertainty should be increased by 18\%, $\delta_{C1} = +1.8 / -1.5$\%.
\\
\\
\noindent
\textbf{ii. Counting efficiency in peaks, $\delta_{C2}$}

The peak counting efficiency is the ratio of the number of registered pulses falling inside the $\pm$FWHM of the energy range of K and L peaks of $^{71}$Ge to the total number of counts. The value was determined by additional measurement using $^{71}$Ge for each counter.
The details of the measurement are described in~Appx.~\ref{app:ge71}.
The average uncertainty for peak efficiencies was $\delta_{C2}\pm1.1$\%.
Considering that four counters were used twice in the BEST measurements this counting systematic uncertainty should be increased by 18\%, $\delta_{C2} = \pm1.3$\%.
\\
\\
\noindent
\textbf{iii. Monte Carlo Interpolation, $\delta_{C3}$}

The counting efficiency is closely related to the pressure~($P$) and the percentage~($G$) of germane in the gas mixture~\cite{sage1999solar}.
The connection was obtained in measurements of volume efficiencies in a wide range of parameters for different counters. 
To obtain the counting efficiencies for arbitrary parameters of the gas mixture, Monte Carlo calculations are used. 
Varying the expression obtained in the analysis of measurements in accordance with the uncertainties of the pressure and the percentage of germane ($\delta P = \delta G = 2$\%), we obtain $\delta_{C3} = \pm 0.6$\% for the K peak and $ \pm 0.4$\% for the L peak.
\\
\\
\noindent
\textbf{iv. Gain variation, $\delta_{C4}$}

During long-term measurements, the gain of the counting system may drift resulting in uncertainty in energy determination. 
The drifts in the positions of the $^{55}$Fe calibration peak were measured during the first month of each counting.  
The average percentage of deviation from the mean values was $\pm 1.1$\%. 
To measure the uncertainty due to the energy determination, the calibration peak positions were shifted by $\pm 1.1$\% and the capture rates in the inner and the outer targets were obtained. 
This variation study yielded the gain variation uncertainty of $\delta_{C4}=+0.4$\%. This uncertainty is one-sided because the gain drift can only reduce the number of selected events.
\\
\\
\noindent
\textbf{v. Resolution and centroid, $\delta_{C5}$ and $\delta_{C6}$}

The systematic uncertainty of the energy resolution of the calibrations was determined similarly to item iv. 
The average percentage of deviation of the resolution of the calibration peaks in the first month of measurements was $\pm$ 2\%, which gives the value $\delta_{C5}=\pm0.3$\%.
Similarly, the deviation of the centroid of the calibration peaks gives the value $\delta_{C6}=\pm0.1$\%.
\\
\\
\noindent
\textbf{vi. Rise time cut, $\delta_{C7}$}

At the end of the measurement, an additional study using $^{71}$Ge was carried out to determine the rise time ($T_N$) values of pulses inside the  2~FWHM regions of the L and K peaks of $^{71}$Ge.
Therefore, in contrast to the previous SAGE experiments where the average $T_N$ limits were used~\cite{sage1999solar}, there is no uncertainty associated with the incorrect determination of the $T_N$ limit interval.
Hence, we assign negligible uncertainty to $\delta_{C7}$.

\subsubsection{Background Discrimination}

Uncertainties arising from background discrimination constitute the third group of uncertainties, $\delta_N$.
\\
\\
\noindent
\textbf{i. Residual radon after the time cuts,  $\delta_{N1}$}

The number of false $^{71}$Ge events due to internal $^{222}$Rn is equal to the number of detected saturated pulses induced by Rn decay~($N_\textrm{Rn}$)  multiplied by the survival probability of false events after the event selection~$\alpha_{K+L}$~\cite{sage1999solar}.
The fraction of the false events to the total number of events $N$ in K and L peaks is treated as a systematic uncertainty

\begin{equation}
\begin{split}
    \delta_{N1} &= \frac{N_\textrm{Rn}\times \alpha_{K+L}}{N} \\
                &= \frac{425.9\times(4.3+7.8)\times10^{-4}}{738.8+724.0} \\
                &= 0.035\%~.
\end{split}
\end{equation}

\noindent
This uncertainty is one-sided negative because the effect of the residual Rn is a reduction in the detected number of events.
\\
\\
\noindent
\textbf{ii. Solar neutrinos,  $\delta_{N2}$}
Solar neutrinos are the largest source of background in the BEST experiment. 
They interact with the Ga targets at a constant rate, and this background is taken into account in the analysis of each measurement.
During solar neutrino measurements, the number of solar-neutrino-induced events in the inner and outer target zones was obtained to be 6.1~(see Table~\ref{tab:inner_measured}) and 32.2~(see Table~\ref{tab:outer_measured}).
These values are compared with the total number of pulses recorded in K and L peaks - 724.0 and 738.8. 
The solar neutrino flux and thus the uncertainty associated with it are obtained from previous gallium solar measurements as ($66.1\pm3.1$)~SNU~\cite{sage1999solar}, the uncertainties are calculated to be $\delta_{N2} = 6.1/724.0\times 3.1/66.1 = 0.04$\% for the inner zone and $\delta_{N2} = 32.2/738.8\times 3.1/66.1 = 0.2$\% for the outer zone. 
\\
\\
\noindent
\textbf{iii. $^{71}$Ge carryover,  $\delta_{N3}$}

The uncertainty associated with the $^{71}$Ge remaining after extractions is found as the product of the statistical uncertainty and the ratio of the carryover events ($N_{CO}$) to the total number of recorded events ($N$)

\begin{equation}
\begin{split}
    \delta_{N3} &= \frac{N_{CO}}{N} \times \delta_N~,
\end{split}
\end{equation}

\noindent
where $\delta_N = 1/\sqrt{N}$~\cite{sage1999solar}.
We obtain 0.03\% and 0.05\% uncertainties for the inner and the outer zones. 

\subsubsection{Average Path Length Uncertainty}

The mass of the internal target obtained from the considered geometry is 63.1~kg more gallium than compared with the mass measurement. 
Such a difference in the target gallium mass will have the maximum effect on the change in the path length if we assume a change in the inner radius of the sphere by 0.18~cm. 
We take the resulting change in the radius as the radius uncertainty, which leads to a systematic uncertainty in the neutrino path length in the inner target of 0.3\%. 
For the external target, the corresponding uncertainty is 0.

The outer radius of the spherical vessel was measured in a direct way, with a ruler, when the sphere vessel was available for measurements in the underground laboratory of the GGNT. 
The surface of the sphere was covered with a layer of epoxy resin. 
The maximum irregularities of the coating did not exceed 1~mm, which varies the average neutrino path length in the outer target of the cylindrical zone by 0.2\%. 
Variations in the internal dimensions of the cylindrical zone, as well as possible inaccuracies in the relative position of the axes of the two zones of the target do not exceed 1~mm. 
A change in the dimensions of the cylindrical zone within the indicated values leads to a change in the average neutrino path length by approximately the same 0.18 cm.
Therefore, we take the total uncertainty of the average neutrino path length in the cylindrical zone to be 0.3\%, {\it i.e.}, the uncertainties of the average neutrino path length in both target zones are assumed to be equal. 
\\
\\
The total systematic uncertainty, added in quadrature, is $-2.7 / +2.9$\%. 
Including both the statistical and systematic uncertainties, we obtain $^{71}$Ge production rates of

\begin{equation}
\begin{split}
    r_\mathrm{In-meas.} &= \left [54.9 \pm2.5(\mathrm{stat.}) ^{+1.6}_{-1.5}(\mathrm{syst.}) \right ] \mathrm {atoms/d}\\
                        &= \left [54.9^{+3.0}_{-2.9}\right ]~\mathrm{atoms/d}~, \\
    r_\mathrm{Out-meas.} &= \left [55.6\pm2.7(\mathrm{stat.}) ^{+1.6}_{-1.5}(\mathrm{syst.})\right ] \mathrm {atoms/d}\\
                         &= \left [55.6^{+3.1}_{-3.1}\right ] \mathrm{atoms/d}~. 
\end{split}
\end{equation}

\begin{figure}[!t]
    \centering
    \includegraphics[width=1.0\columnwidth]{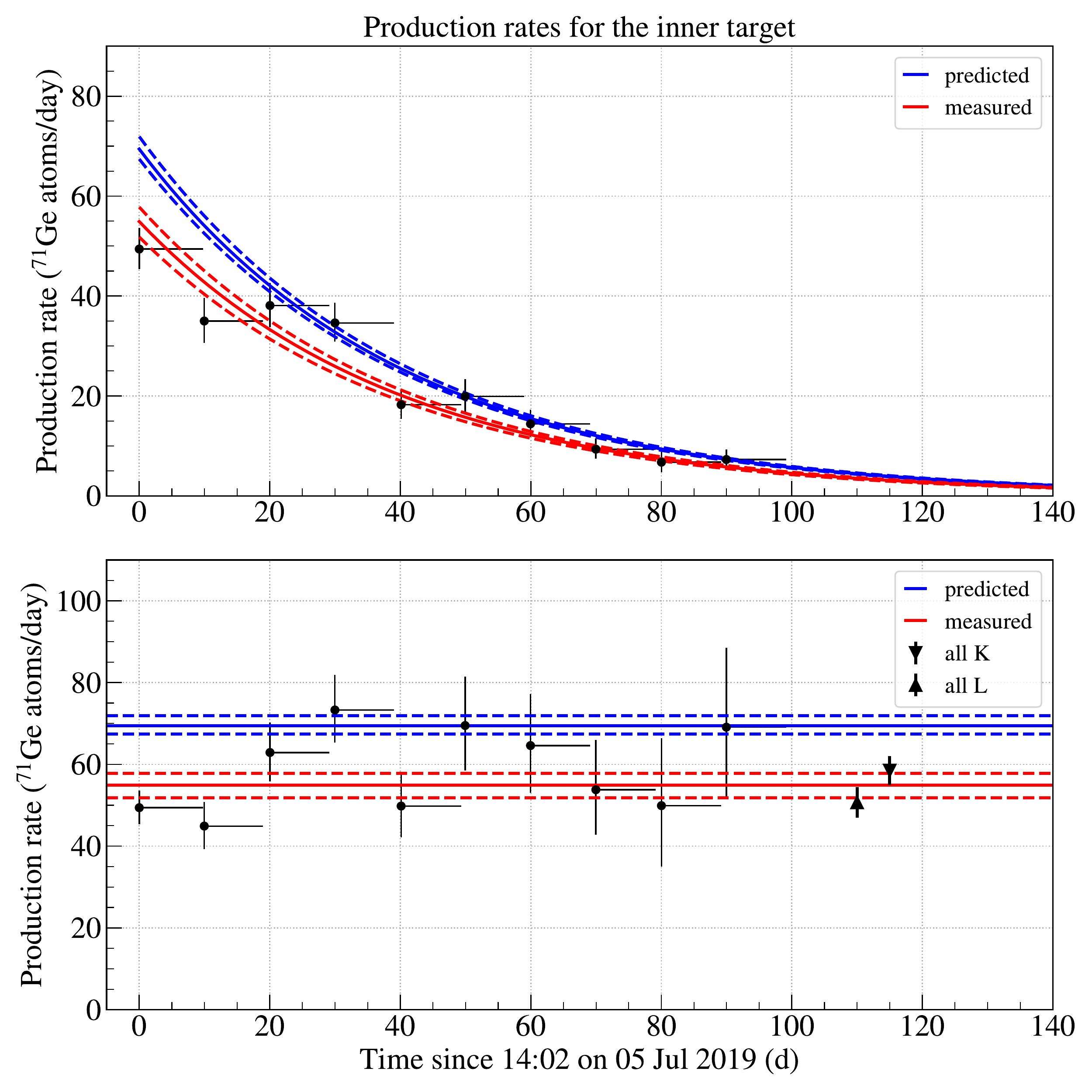}
    \includegraphics[width=1.0\columnwidth]{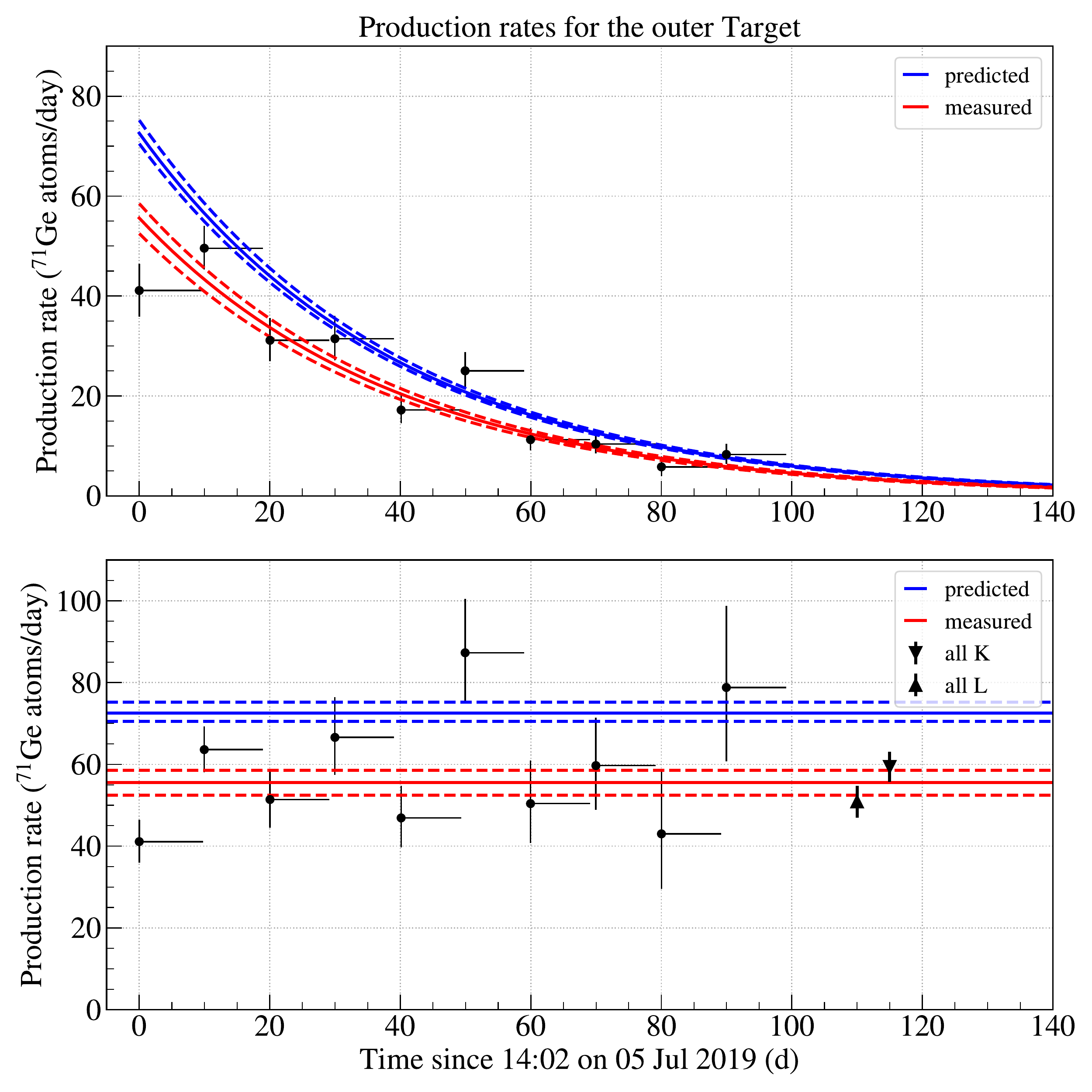}
    \caption{\textbf{Top:} the measured K+L peak rates of the inner target volume. \textbf{Middle-top:} the production rates of the inner target volume normalized to the reference time. The combined results for events in the L and K peaks are shown separately and compared to the predicted rate.  \textbf{Middle-bottom:} the measured K+L peak rates of the outer target volume. \textbf{Bottom:} the production rates of the outer target volume normalized to the reference time. The combined results for events in the L and K peaks are shown separately and compared to the predicted rate.  The blue (red) region represents the predicted (measured) production rate. The dotted lines enclose the $\pm 1 \sigma$ uncertainty regions.}
    \label{fig:production_rate}
\end{figure}

\begin{table*}
    \caption{Summary of BEST detection efficiencies. The extraction efficiency values are taken from Tables~\ref{tab:inner_extraction} and~\ref{tab:outer_extraction}. The counter efficiency values and $\Delta$ are taken from Tables~\ref{tab:inner_counting} and \ref{tab:outer_counting}. Same fractional uncertainties listed in Table~\ref{tab:systematic_uncertainty} are assigned to each extraction. The saturation factor is obtained from Eq.~\ref{eq:saturation_factor} with negligible uncertainty. }
    \centering
    \begin{tabular}{|l||c|c||c|c|c||c|c|c|}
    \hline
    Extraction	& \makecell{Extraction  \\ efficiency \\ (into GeH$_4$)} & \makecell{Saturation \\ factor} & \makecell{Counter \\ efficiency \\ (K peak)} &  $\Delta_\mathrm{K}$ & \makecell{Total detection \\ efficiency\\ (K peak)} & \makecell{Counter \\ efficiency \\(L peak)} & $\Delta_\mathrm{L}$ &  \makecell{Total detection \\efficiency \\ (L peak)} \\
    \hline
Inner-1	&  0.946	&  0.3924	&  0.3663	&  0.8100	&  0.1101	&  0.3803	&  0.7450		&  0.1052 \\
Inner-2	&  0.9559	&  0.3731	&  0.3647	&  0.7839	&  0.1020	&  0.3785	&  0.6542		&  0.0883 \\
Inner-3	&  0.9673	&  0.3752	&  0.3605	&  0.7143	&  0.0935	&  0.3599	&  0.7143		&  0.0933 \\
Inner-4	&  0.9515	&  0.3750	&  0.3679	&  0.7870	&  0.1033	&  0.3769	&  0.3669		&  0.0493 \\
Inner-5	&  0.9554	&  0.3779	&  0.3649	&  0.7470	&  0.0984	&  0.3654	&  0.7470		&  0.0986 \\
Inner-6	&  0.9548	&  0.3735	&  0.3577	&  0.7714	&  0.0984	&  0.3604	&  0.3891		&  0.0500 \\
Inner-7	&  0.9381	&  0.3760	&  0.3676	&  0.7495	&  0.0972	&  0.3793	&  0.6739		&  0.0902 \\
Inner-8	&  0.9789	&  0.3761	&  0.3656	&  0.7754	&  0.1044	&  0.3779	&  0.7865		&  0.1094 \\
Inner-9	&  0.9545	&  0.3761	&  0.359	&  0.8015	&  0.1033	&  0.361	&  0.8015		&  0.1039 \\
Inner-10	&  0.9372	&  0.3770	&  0.3698	&  0.7629	&  0.0997	&  0.3755	&  0.8002		&  0.1062 \\
\hline
Outer-1	&  0.9503	&  0.3924	&  0.3422	&  0.7646	&  0.0976	&  0.3596	&  0.6996		&  0.0938 \\
Outer-2	&  0.9581	&  0.3731	&  0.3707	&  0.8043	&  0.1066	&  0.3792	&  0.6755		&  0.0916 \\
Outer-3	&  0.9668	&  0.3752	&  0.2933	&  0.7650	&  0.0814	&  0.3358	&  0.7650		&  0.0932 \\
Outer-4	&  0.9622	&  0.3750	&  0.3658	&  0.7819	&  0.1032	&  0.381	&  0.3755		&  0.0516 \\
Outer-5	&  0.9609	&  0.3779	&  0.3568	&  0.8025	&  0.1040	&  0.3727	&  0.8025		&  0.1086 \\
Outer-6	&  0.9253	&  0.3735	&  0.3585	&  0.8009	&  0.0992	&  0.3577	&  0.3845		&  0.0475 \\
Outer-7	&  0.9514	&  0.3760	&  0.3407	&  0.7976	&  0.0972	&  0.3607	&  0.7107		&  0.0917 \\
Outer-8	&  0.9897	&  0.3761	&  0.3716	&  0.8295	&  0.1147	&  0.3785	&  0.8406		&  0.1184 \\
Outer-9	&  0.9664	&  0.3761	&  0.293	&  0.7865	&  0.0838	&  0.336	&  0.7865		&  0.0961 \\
Outer-10	&  0.9538	&  0.3770	&  0.3677	&  0.7567	&  0.1001	&  0.3797	&  0.7940		& 0.1084 \\
    \hline
    \end{tabular}
    \label{tab:systematic_efficiency}
\end{table*}

Various efficiencies for individual extraction are collected and summarized in Table~\ref{tab:systematic_efficiency}. 
The fractional uncertainties are the same for every extractions and given in Table~\ref{tab:systematic_uncertainty}. 
The saturation factor in the table represents the saturation in production rate due to $^{71}$Ge and $^{51}$Cr activities and is given by 

\begin{equation}\label{eq:saturation_factor}
\begin{split}
    f_s =&~\mathrm{exp}(-\lambda_{51}(t_s-T)) \\
    & \times[\mathrm{exp}(-\lambda_{51}\theta_\mathrm{Cr})-\mathrm{exp}(-\lambda_{71}\theta_\mathrm{Cr})]/(1-\lambda_{51}/\lambda_{71})~,  
\end{split}
\end{equation}

\noindent
where $\lambda_{51}$ and $\lambda_{71}$ are the decay constants of $^{51}$Cr and $^{71}$Ge, $t_s$ is the start time of each source exposure, $T$ is the source activity reference time and $\theta_\mathrm{Cr}$ is the time of exposure of the Ga to the $^{51}$Cr source~\cite{sage1999source}\footnote{One should note that there is a typographical error in Eq.~4 of Ref.~\cite{sage1999source}. There is a missing division sign in the denominator, which is corrected in Eq.~\ref{eq:saturation_factor} here in this work. The correct denominator is $(1 - \lambda_{51}/\lambda_{71})$.}.
All parameters considered in the calculation of $f_s$ are very well established, and therefore its uncertainty is negligible.
The total efficiency is the product of the chemical extraction efficiency, the saturation factor, and the counting efficiency~\cite{sage1999source}.
The typical value of the total efficiency is (10.0$\pm$0.3\%), where systematic uncertainties are included.

Fig.~\ref{fig:production_rate} shows the K+L production rate fits for the inner and the outer zones. The upper panel for each zone shows the production rate at the start of each exposure, and the lower panel shows the rate normalized to the reference time.
The results are plotted at the start of each exposure where the production rates are fit to.
The vertical lines represent the uncertainty of the fits, while horizontal lines only define the exposure period.
The ratio of measured to predicted production rates are

\begin{equation}
\begin{split}
    R&_\mathrm{In} = \frac{54.9^{+3.0}_{-2.9}}{69.4^{+2.5}_{-2.0}} = 0.79^{+0.05}_{-0.05}~,\\
    R&_\mathrm{Out} = \frac{55.6^{+3.1}_{-3.1}}{72.6^{+2.6}_{-2.1}} = 0.77^{+0.05}_{-0.05}~.
\end{split}
\end{equation}

\noindent
These are 4.2 $\sigma$ and 4.8 $\sigma$ less than unity, respectively.
As a cross-check, the $^{51}$Cr half-life was left free and fitted to the data. 
The values of half-life from fits to the inner and the outer targets were ($30.97\pm3.90$)\,d and ($31.55\pm2.89$)\,d. 
They agree with the known $^{51}$Cr half-life of (27.704$\pm$0.004)~\cite{Zhou1991}. 

The ratio between the inner and the outer zones ${R}_\textrm{out}/{R}_\textrm{in} = (0.77\pm0.05)/(0.79\pm0.05) = 0.97\pm0.08$  is unity within uncertainty, and hence there is no difference in the capture rates between the two zones.

\begin{figure}[!t]
    \centering
    \includegraphics[width=\columnwidth]{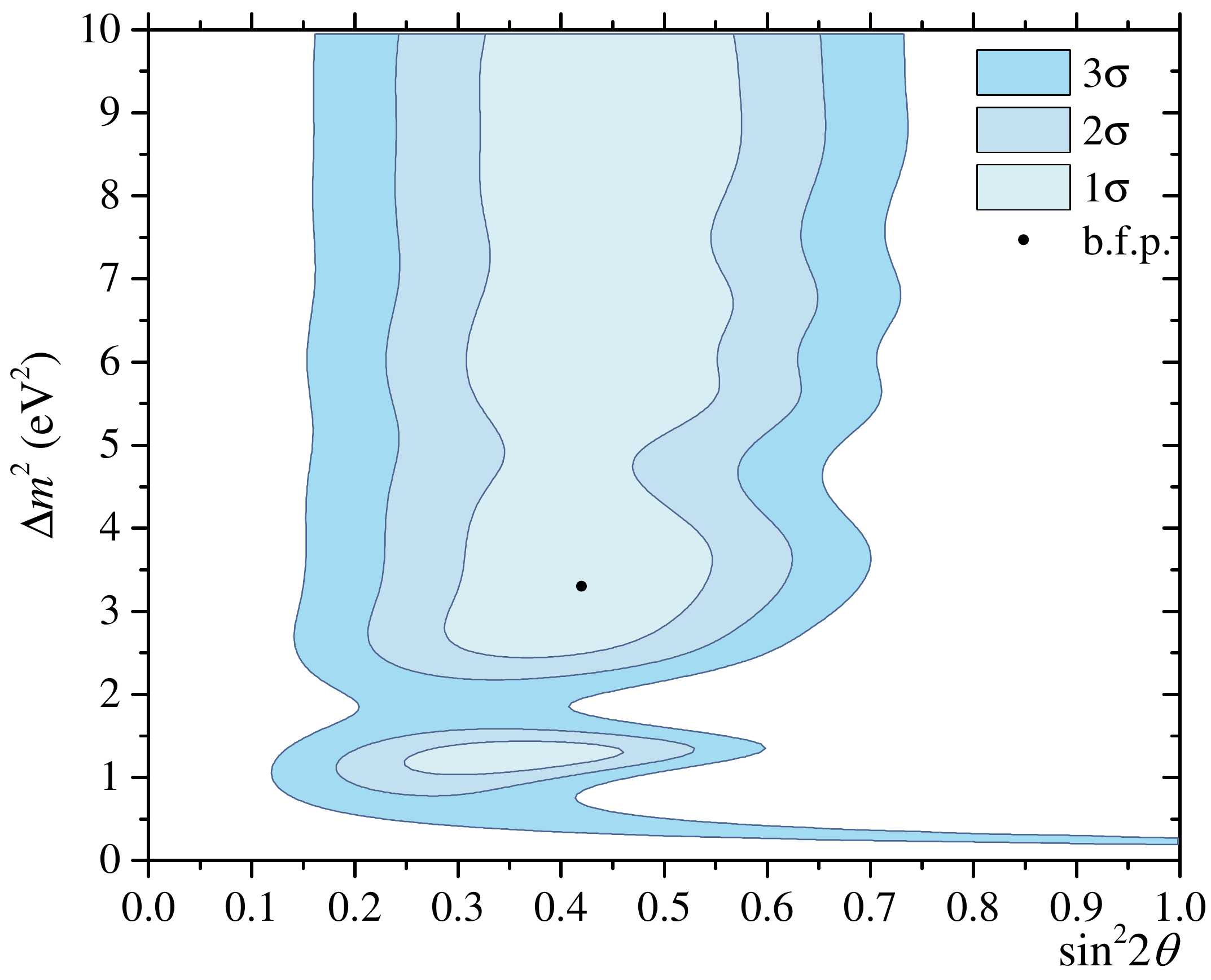}
    \caption{Allowed regions for two BEST results. The best-fit point is  sin$^2 2\theta=0.42^{+0.15}_{-0.17}$,  $\Delta m^2=3.3^{+\infty}_{-2.3}$~eV$^2$ and is indicated by a point.}
    \label{fig:best_contour}
\end{figure}

\begin{figure}[!t]
    \centering
    \includegraphics[width=\columnwidth]{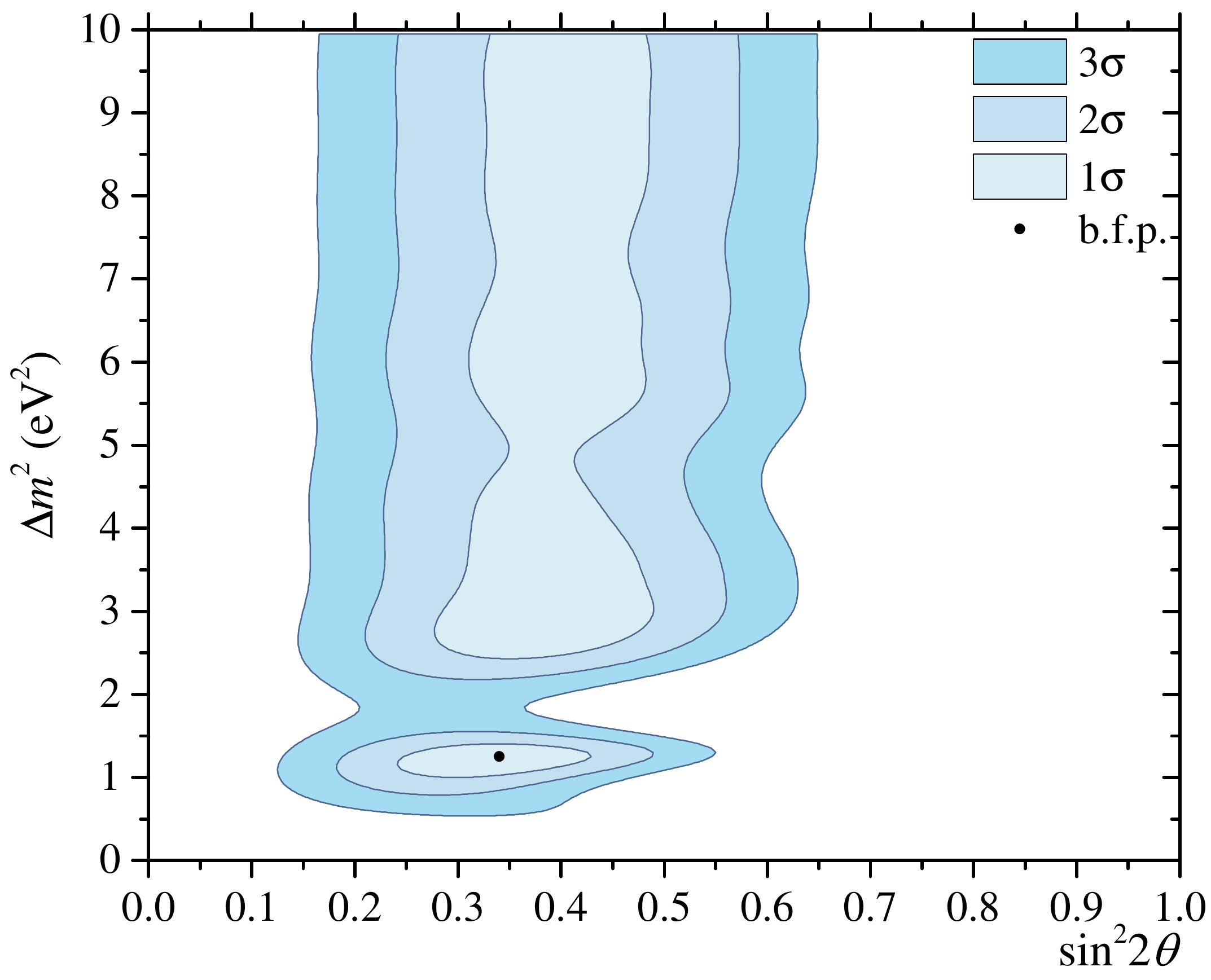}
    \caption{Allowed regions for two GALLEX, two SAGE and two BEST results. The best-fit point is sin$^2 2\theta=0.34^{+0.14}_{-0.09}$, $\Delta m^2 = 1.25^{+\infty}_{-0.25}$ eV$^2$ and is indicated by a point.}
    \label{fig:combined_contour}
\end{figure}

\subsection{Oscillation Analysis}

For $n=1,\dots,N$ experiments (the two BEST volumes are treated separately), oscillation parameters are estimated by a global minimization of 
\begin{equation}\label{eqn:chisquare}
\small
\chi^2(\Delta m^2,\sin^22\theta)=({\bf {r}^\mathrm{meas.}-{r}^\mathrm{calc.}})^\mathrm{T}{\bf V}^{-1} ({\bf {r}^\mathrm{meas.}-{r}^\mathrm{calc.}})~,
\end{equation}
where ${\bf {r}^\mathrm{meas.}}$ (${\bf {r}^\mathrm{calc.}}$) is the vector of the measured (calculated) rates with ${r}_i^\mathrm{calc.}(\Delta m^2,\mathrm{sin}^2 2\theta)$ and $V$ is the covariance matrix with
\begin{equation}
V_{nk}=\delta_{nk}\varepsilon_n^2 + \varepsilon_{CS}^n \times \varepsilon_{CS}^k~,
\end{equation}
where $\varepsilon_n^2=\varepsilon_{n,stat}^2+\varepsilon_{n,syst}^2$ are uncorrelated uncertainties comprised of statistical and systematic measurement uncertainties, and  $\varepsilon_{CS}^n$ represent the correlated uncertainties of $\sigma$~\cite{Fogli2002}.
For Ga source experiments, the cross-section uncertainties are the only significant contribution to the correlated uncertainty. 
The uncertainties of the Bahcall cross-section for measurements with a chromium source are equal to +3.6/–2.8\%, and hence we use $\pm3.6$\% for the calculation.
In the measurement with an argon source, the cross-section uncertainties do not change the values of the shown uncertainty due to their relative smallness.

Fig.~\ref{fig:best_contour} illustrates the exclusion contours corresponding to $1\sigma, 2\sigma$, and $3\sigma$ confidence levels with the two BEST results assuming  the correlated cross section uncertainties.
The best-fit result is at sin$^2 2\theta=0.42^{+0.15}_{-0.17}$,  $\Delta m^2=3.3^{+\infty}_{-2.3}$~eV$^2$, and the contours refer to $\chi^2 = \chi^2_\textrm{min} + \Delta \chi^2$ for which $\Delta \chi^2 = 2.30, 6.18, 11.83$ with two degrees of freedom have the coverage probability of 68.27\% (1$\sigma$), 95.45\% (2$\sigma$) and 99.73\% (3$\sigma$), respectively~\cite{Olive_2014}.
This approximation is based on Wilks' theorem.
A concern about the applicability of Wilks' theorem in the oscillation analysis~\cite{coloma2021statistical} was investigated and found to have a small effect on the analysis~\cite{berryman2021statistical}.

\begin{table}[t]
    \caption{Results of all six Ga source experiments.}
    \centering
    \begin{tabular}{|l|c|}
    \hline
         Experiment & $R$  \\
         \hline
         SAGE-Cr~\cite{sage1999source}  & 0.95 $\pm$ 0.12 \\
         SAGE-Ar~\cite{sage2006argon}  & 0.79  $\pm$ 0.095 (+0.09 / -0.10) \\
         GALLEX-Cr1~\cite{gallex2010reanalysis}  & 0.953 $\pm$ 0.11 \\
         GALLEX-Cr2~\cite{gallex2010reanalysis}  & 0.812 $\pm$ 0.11 \\
         BEST-Inner  & 0.791 $\pm$ 0.05 \\
         BEST-Outer  & 0.766 $\pm$ 0.05 \\
         \hline
    \end{tabular}
    \label{tab:gallium_all}
\end{table}

\begin{figure}[t]
    \centering
    \includegraphics[width=\columnwidth]{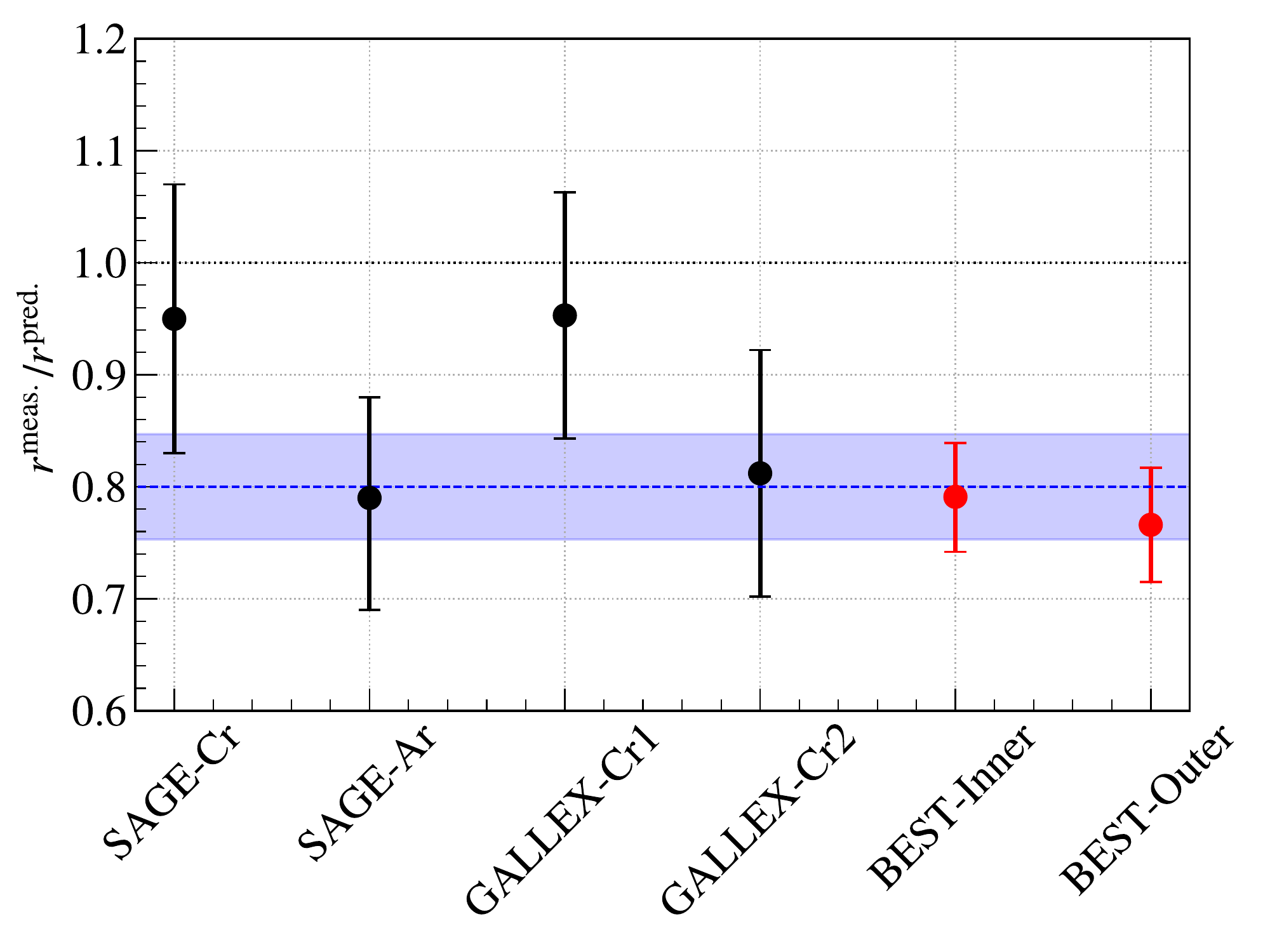}
    \caption{Ratios of measured and predicted $^{71}$Ge production rates in all Ga source experiments. The combined result is shown as a blue band.}
    \label{fig:production_rates_all}
\end{figure}

We also combine the BEST results with the previous SAGE and GALLEX source experiments to plot combined exclusion plots.
Table~\ref{tab:gallium_all} and Fig.~\ref{fig:production_rates_all} summarize results from all gallium anomaly experiments, including the SAGE~\cite{sage1999source,sage2006argon} and the GALLEX results~\cite{gallex2010reanalysis}.

If the cross-section uncertainties are considered to be completely uncorrelated to each other and to the other uncertainties, they can be added in quadrature as

\begin{equation}
    \sigma_i = \sqrt{\sigma_{i,\mathrm{others}}^2 + (0.032\times R_i)^2}~,
\end{equation}

\noindent
and the combined result $R_0$ is obtained by the sum $R_0 = \sum_i(w_i\cdot R_i)$, where $w_i = (\sigma_0/\sigma_i)^2$ and $\sigma_0 = 1/\sqrt{\sum_i (1/\sigma_i^2)}$.
The result is given as $R_0\pm\sigma_0 = 0.81 \pm 0.03$. The total uncertainty is 4.0\%.

If we consider the correlation between systematic uncertainties, the average value of $R_{0,\mathrm{Cr}}$ is obtained first, and then combined with the SAGE-Ar experiment afterward. The uncertainty from the cross section evaluation is the only significant contribution to the correlated uncertainty, and hence the combined result of all six gallium anomaly experiments is given as

\begin{equation}
    R_0 = \Big(\frac{\sigma_R}{\sigma_\mathrm{Cr}}\Big)^2 \cdot R_\mathrm{Cr} + \Big(\frac{\sigma_R}{\sigma_\mathrm{Ar}}\Big)^2 \cdot R_\mathrm{Ar}\pm\sigma_R,
\end{equation}

\noindent
where $\sigma_R =  1/\sqrt{(1/\sigma_\mathrm{Cr}^2+1/\sigma_\mathrm{Ar}^2)}$, $R_\mathrm{Cr} = \sum_i^\mathrm{Cr}(w_i\cdot R_i)$, $w_i = (\sigma_0/\sigma_{i,\mathrm{others}})^2$, $\sigma_0 =  1/\sqrt{\sum_i^\mathrm{Cr} (1/\sigma_{i,\mathrm{others}}^2)}$ and $\sigma_\mathrm{Cr} = \sigma_0 + 0.032\cdot R_\mathrm{Cr}$, with the uncertainty of the total Cr measurements is obtained by summing over. 
The combined result of all six measurements is obtained as $R_0 = 0.80\pm0.047$. The total uncertainty is 6.1\%, which is larger than the uncorrelated estimation. 
The result is illustrated in  Fig.~\ref{fig:production_rates_all} as a blue band.

Fig.~\ref{fig:combined_contour} presents the combined result from all gallium source experiments; SAGE, GALLEX and BEST, considering the correlated cross section uncertainties.
The best-fit result from the combined analysis of all Ga source experiments is  sin$^2 2\theta=0.34^{+0.14}_{-0.09}$, $\Delta m^2 = 1.25^{+\infty}_{-0.25}$ eV$^2$.

\begin{figure}
    \centering
    \includegraphics[width=1.0\columnwidth]{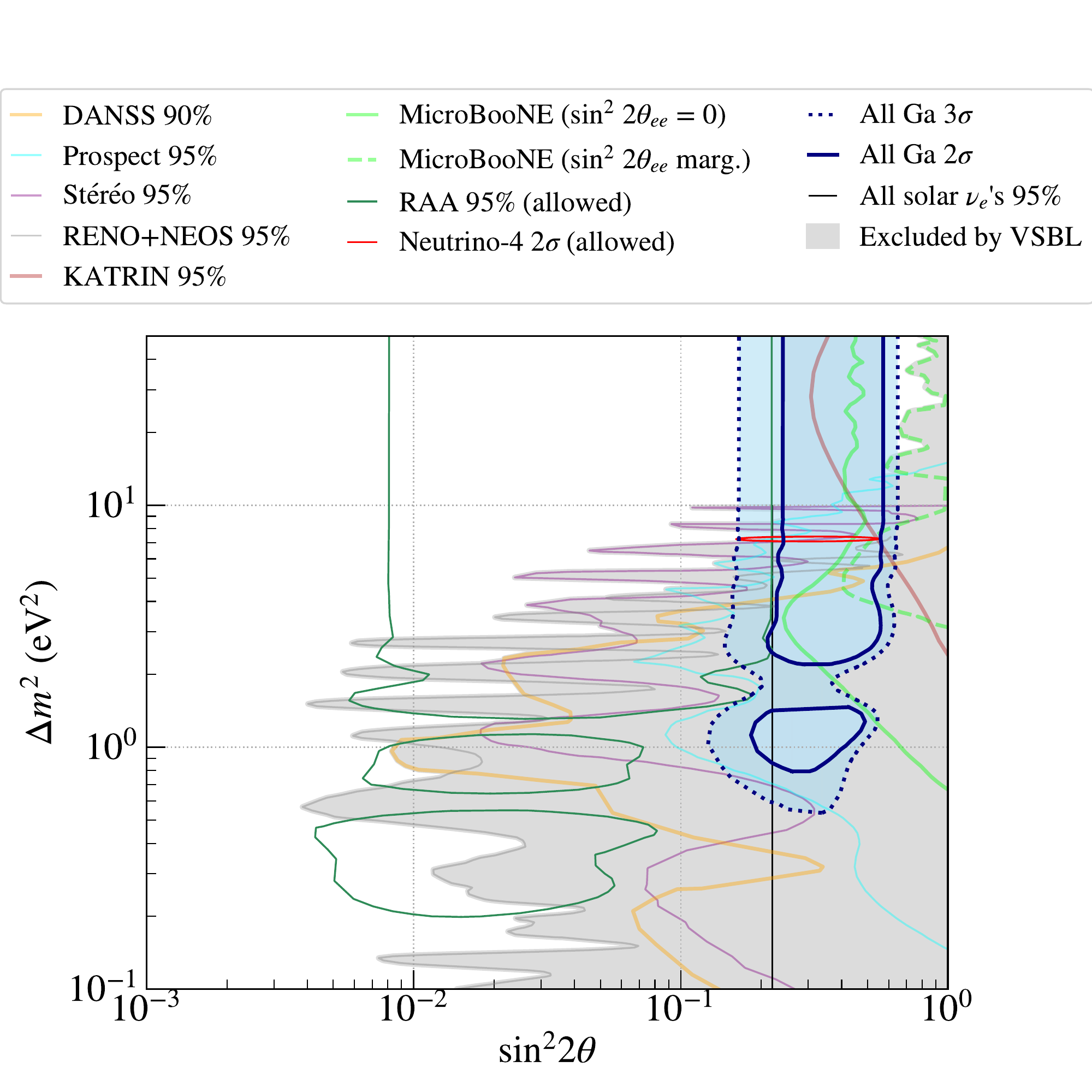}
\caption{Exclusion contours of all gallium anomaly experiments: two GALLEX, two SAGE and two BEST results. The blue solid line and the blue dotted line show the 2$\sigma$ and 3$\sigma$ confidence level, respectively. The figure also presents the exclusion contours from Prospect~\cite{prospect2021}, DANSS~\cite{danss2020}, St\'{e}r\'{e}o~\cite{stereo2020}, KATRIN~\cite{katrin2020}, the combined analysis of RENO and NEOS data~\cite{atif2020search}, reactor anti-neutrino anomalies~(RAA)~\cite{chooz2011reactor}, interpretations of the MicroBooNE result for the oscillation hypothesis with fixed mixing angle~(sin$^22\theta$) and profiled over the angle~\cite{arguelles2021microboone}, and the model-independent 95\% upper bound on sin$^2 2\theta$ from all solar neutrino experiments~\cite{neutrino4_2021}. The 2$\sigma$ allowed region of Neutrino-4~\cite{serebrov2020analysis} is also presented and the grey shading represents the merged exclusion of the very short baseline~(VSBL) null results.}
    \label{fig:all_combined}
\end{figure}

Fig.~\ref{fig:all_combined} compares the combined result from all gallium anomaly experiments to some other sterile neutrino search experiments. 
The exclusion curves of Prospect~\cite{prospect2021}, DANSS~\cite{danss2020}, St\'{e}r\'{e}o~\cite{stereo2020}, KATRIN~\cite{katrin2020}, the combined analysis of RENO and NEOS data~\cite{atif2020search}. 
One can see that the gallium anomaly result is still in a strong tension with these experiments except a tiny region above 8~eV$^{2}$.
The interpretations of the MicroBooNE result for the oscillation hypothesis either fixed or  profiled over the mixing angle~(sin$^22\theta$)~\cite{arguelles2021microboone} are also presented.
These results do not either favor or exclude the allowed region for the gallium anomaly experiments.
The 95\% allowed region from the reactor antineutrino anomalies (RAA)~\cite{chooz2011reactor} is also illustrated in the figure. 
One can see that the tension between the Ga anomalies and the RAA still persists.
The figure also shows the 2$\sigma$ allowed region of Neutrino-4~\cite{serebrov2020analysis} and the model-independent 95\% solar neutrino upper bound on sin$^2 2\theta$~\cite{neutrino4_2021}. 
The Neutrino-4 allowed region is within the 3$\sigma$ bound, and therefore is not excluded by the gallium anomaly experiments. The model-independent 95\% solar neutrino bound on sin$^22\theta$ excludes most of the 2$\sigma$ allowed region, but the parameter space near $\Delta m^2\approx$ 1~eV$^2$ is not entirely excluded.

\section{Discussion} \label{sec:discussion}

\begin{table*}[tp]
    \centering
    \caption{A summary of the $^{71}$Ge counter calibration. }
    \footnotesize
    \begin{tabular}{|l||c|c|c||c|c|c|c|c|c||c|c|c|c|c|c|}
    \hline
    \multirow{3}{*}{Counter}  & \multicolumn{3}{c||}{$^{55}$Fe} &  \multicolumn{6}{c||}{L-peak} &  \multicolumn{6}{c|}{K-peak} \\
    \cline{2-16}
    & & & & \multicolumn{3}{c|}{Position} & \multicolumn{3}{c||}{Resolution} &  \multicolumn{3}{c|}{Position} & \multicolumn{3}{c|}{Resolution} \\
    \cline{5-16}
    & \makecell{Pos. \\ (a.u.)} & \makecell{Resol. \\ (\%)} & \makecell{En \\ non-linearity \\ factor} 
    & \makecell{Pred. \\ from \\$^{55}$Fe \\ (a.u.)} & \makecell{True\\ (a.u.)} &  \makecell{Ratio} 
    & \makecell{Pred. \\ from \\$^{55}$Fe\\ (\%)} & \makecell{True\\ (\%)} &  \makecell{Ratio} 
    & \makecell{Pred. \\ from \\$^{55}$Fe\\ (a.u.)} & \makecell{True\\ (a.u.)} &  \makecell{Ratio} 
    & \makecell{Pred. \\ from \\$^{55}$Fe\\ (\%)} & \makecell{True\\ (\%)} &  \makecell{Ratio} 
    \\
    \hline
    \hline
    \makecell[l]{\textbf{sys2z}\\ } & & & & & & & & & & & & & & & \\
    YCN43 & 437.39 & 18.2 & 0.982 & 86.81 & 87.16 & 1.00 & 40.92 & 40.29 & 0.98 & 755.35 & 725.50 & 0.96 & 15.88 & 12.98 &  0.82 \\
    YCNA9 & 431.86 & 19.7 & 0.938 & 85.71 & 86.32 & 1.01 & 44.29 & 42.57 & 0.96 & 712.39 & 728.84 & 1.02 & 17.19 & 16.02 &  0.93 \\
    YCN41 & 442.80 & 18.0 & 0.967 & 87.88 & 87.00 & 0.99 & 40.40 & 37.24 & 0.92 & 753.01 & 732.49 & 0.97 & 15.68 & 13.34 &  0.85 \\
    YCN46 & 448.93 & 18.5 & 0.915 & 89.10 & 90.61 & 1.02 & 41.48 & 39.62 & 1.04 & 722.39 & 748.55 & 0.96 & 16.10 & 14.16 &  0.88 \\
    \hline
    \makecell[l]{Mean sys2z \\ std} &   &   &  & &  & \makecell{1.00 \\ 0.01} & & & \makecell{0.96 \\ 0.03} & & & \makecell{1.00 \\ 0.04}  & & & \makecell{0.87 \\ 0.05} \\
    \hline
    \makecell[l]{\textbf{sys3}\\ } & & & & & & & & & &  && & & & \\
    YCN113 & 337.36 & 18.8 & 0.982 & 66.96 & 57.71 & 0.86 & 42.09 & 42.13 & 1.00   & 582.61 & 564.38 & 0.97 & 16.34 & 14.13 & 0.86 \\
    YCT92  & 334.67 & 18.9 & 0.964 & 66.42 & 56.72 & 0.85 & 42.40 & 41.88 & 0.99   & 567.37 & 563.97 & 0.99 & 16.46 & 13.90 & 0.84 \\
    YCT3   & 342.65 & 19.3 & 0.932 & 68.01 & 58.24 & 0.86 & 43.23 & 40.61 & 0.94   & 561.61 & 576.97 & 1.03 & 16.78 & 14.19 & 0.85 \\
    YCT2   & 337.00 & 18.8 & 0.982 & 66.89 & 57.77 & 0.86 & 42.24 & 41.62 & 0.99   & 581.98 & 563.33 & 0.97 & 16.40 & 14.05 & 0.86 \\
    YCT9   & 334.80 & 19.1 & 0.978 & 66.45 & 57.97 & 0.87 & 42.90 & 39.42 & 0.92   & 575.83 & 569.62 & 0.99 & 16.65 & 13.89 & 0.83 \\
    YCT97  & 332.81 & 18.6 & 0.932 & 66.05 & 57.11 & 0.86 & 41.64 & 38.75 & 0.93   & 545.48 & 559.25 & 1.03 & 16.16 & 13.51 & 0.84 \\
    \hline
    \makecell[l]{Mean sys3 \\ std} &   &   &   & & &  \makecell{0.86 \\ 0.01} && &  \makecell{0.96 \\ 0.03} & & & \makecell{1.00 \\ 0.03}  & & & \makecell{0.85 \\ 0.01} \\
    \hline
    \hline
    \end{tabular}
    \label{tab:counter_calibration}
\end{table*}

The deficits in the production rates observed in both volumes are consistent with the results from previous gallium source experiments, confirming the gallium anomaly. 
While no difference in the production rates in the two target zones is observed and the oscillation length cannot be well constrained, the large deviation of R's from unity indicates that the oscillation hypothesis is favored over no-oscillation hypothesis to the extent of current knowledge. 
Effects from systematic errors in the extraction efficiency, the counting efficiency, Ga target mass, the geometry of the system, the source strength and the counting system operation have been thoroughly considered and confirmed by additional measurements to be small~(see Tables~\ref{tab:predict_calculation}~and~\ref{tab:systematic_uncertainty}) compared to the observed deficits. 
In addition, GALLEX's $^{71}$As experiment~\cite{gallex1998arsenic} ruled out any chemical effects of 'hot atoms' that would make it difficult to extract $^{71}$Ge atoms formed during neutrino capture.
Since the physics processes involved in the Ga experiments are simple and understood very well, it is hard to attribute the result to some unaccounted effects. 
The $^{71}$Ge decay rate is very well known, and the neutrino-nucleon cross-section $\sigma$ to the ground state in $^{71}$Ga is hence also well determined. The $\sigma$ cannot be reduced below the ground state value by considering the excited state contribution to be zero. 
The transition to excited states cannot decrease the overall value of the total $\sigma$ and hence cannot be the origin of the reported deficits. 
When the solar neutrino flux measured by the SAGE~\cite{abdurashitov2009measurement} is compared to the Borexino~\cite{agostini2018comprehensive} result, the two experiments claim similar $pp$ flux. SAGE measured the $pp$ flux produced in the Sun to be (6.0$\pm$0.84)$\times10^{10}$ / (cm$^2$~s), while Borexino yielded $6.1\pm0.05^{+0.3}_{-0.5}\times10^{10}$ / (cm$^2$~s). However, the uncertainties are too large to either confirm or exclude the overall efficiency as explanation of the gallium anomaly.
Other explanations without sterile neutrinos are possible but would inevitably require some fundamental misunderstanding of nuclear or atomic physics.

After the BEST measurements the gallium anomaly is more significant; the weighted average value of the neutrino capture rate relative to the expected value for all Ga experiments is $0.80\pm 0.05  $.
The result indicates that the oscillation length is similar to, or smaller than, the volume of the BEST targets. 
Hence, $\Delta m^2$ is unbounded from above by the BEST experiment. 
A future Ga experiment with shorter baseline would help constraining the value of $\Delta m^2$, but it would require the use of stronger neutrino source which would be challenging. The use of neutrino source with larger neutrino energy could also be considered.

\section{Conclusion} \label{sec:conclusion}

The first result from the BEST sterile neutrino oscillation experiment was presented. 
We report 21\% and 23\% deficits of the $^{71}$Ge production rates based on the cross section~\cite{Bahcall1997} in the inner and the outer zones of the BEST Ga target. 
The values are consistent with the previously reported gallium anomaly from SAGE and GALLEX experiments, with higher significance. 
The weighted average value of the neutrino capture rate relative to the expected value for all Ga experiments is $0.80\pm 0.05$, which corresponds to a 4$\sigma$ deviation from unity.
If attributed to the neutrino oscillation to sterile state, the result corresponds to the best fit of $\Delta m^2=3.3^{+\infty}_{-2.3}$~eV$^2$ and sin$^2 2\theta=0.42^{+0.15}_{-0.17}$.
No difference in the capture rates from the two zones at different distances is observed. 
This indicates a neutrino oscillation at a scale shorter than the BEST dimensions. 

The allowed region in the ($\Delta m^2$, sin$^22\theta$) parameter space for all combined Ga source experiments shows that the Ga anomalies are still in tension with most other sterile neutrino search experiments. 
On the other hand, the Neutrino-4 allowed region coincides with the gallium anomaly result, and therefore is not rejected by the Ga source experiments. 
The value of $\Delta m^2$ should be better constrained to conclusively determine whether the Neutrino-4 result is accepted or rejected by the Ga source experiments.

Ga source experiments are unique in their capability of searching for the very short range neutrino oscillation at a meter scale, and the simplicity of the physics involved. 
They can provide critical information on the field of the sterile neutrino search experiments, providing a key to detect the neutrino oscillation to sterile states and the opportunity of resolving the existing tension between the reactor antineutrino experiments and other experiments. 
Further study with other neutrino sources will enhance our knowledge on the gallium anomaly, and hence lead to an irreplaceable contribution to the pursuit of the physics beyond the Standard Model.

\begin{acknowledgments}
We thank V. A. Rubakov for constant stimulation of our interest and for fruitful discussions. This work is supported by Federal Agency for Scientific Organizations (FANO), Ministry of Education and Science of Russian Federation under agreement no. 14.619.21.0009 (unique project identifier no. RFMEFI61917X0009), State Atomic Energy Corporation Rosatom, and the Office of Nuclear Physics of the US Department of Energy.
\end{acknowledgments}

\appendix

\begin{table*}[th]
    \caption{A summary of the $T_N$ event selection efficiency test using $^{71}$Ge calibration data.}
    \centering
    \begin{tabular}{|l|c|c|c|c|c|c|c|c|}
    \hline
    Counter	& \makecell{Num. Ev. \\ L- all} & \makecell{L-peak \\ $T_N$ cut (ns)}	& \makecell{Num. Ev \\L- selected} & ratio & 
        	  \makecell{Num. Ev. \\ k- all} & \makecell{K-peak \\ $T_N$ cut (ns)}	& \makecell{Num. Ev \\k- selected} & ratio \\
    \hline
    \multicolumn{9}{|l|}{sys2z} \\
    \hline
    YCN43 & 495 & 10 & 463 & 0.94 & 489 & 13.2 & 468 & 0.96 \\ 
    YCNA9	& 1281	& 13.2	& 1244	& 0.97	& 1167	& 18.8	& 1142	& 0.98 \\
    YCN41	& 1353	& 10.3	& 1299	& 0.96	& 1434	& 13.4	& 1374	& 0.96 \\
    YCN46	& 941	& 11.3	& 897	& 0.95	& 865	& 15.2	& 837	& 0.97 \\
    \hline
    \makecell[l]{mean 2z \\ std}& 	& 	& 		& \makecell{0.95 \\ 0.01} & 	& 	& 		& \makecell{0.97 \\ 0.01} \\
    \hline
    \hline
    \multicolumn{9}{|l|}{sys3} \\
    \hline
    YCN113  & 1643  & 9.1   & 1626  & 0.99  & 1488  & 13.6 & 1426   & 0.96 \\ 
    YCT92	&  265	& 13.0	& 250	& 0.94	& 243	& 17.6	& 237	& 0.98 \\ 
    YCT4	& 508	& 10.2	& 497	& 0.98	& 328	& 13.2	& 313	& 0.95 \\
    YCT3	& 314	& 10.3	& 297	& 0.95	& 258	& 16.4	& 252	& 0.98 \\
    YCT2	& 1475	& 10.1	& 1415	& 0.96	& 1483	& 16.6	& 1427	& 0.96 \\
    YCT9	& 397	& 9.1	& 388	& 0.98	& 341	& 14.9	& 322	& 0.94 \\
    YCT97	& 1622	& 11.4	& 1551	& 0.96	& 1607	& 17.3	& 1562	& 0.97 \\
    \hline
    \makecell[l]{mean sys3 \\ std}& 	& 	& 		& \makecell{0.96 \\ 0.02} & 	& 	& 		& \makecell{0.96 \\ 0.02} \\
    \hline
    \end{tabular}
    \label{tab:tn_test}
\end{table*}

\begin{table*}[t]
    \centering
    \caption{BEST results without the first Cr extractions}
    \begin{tabular}{|l|r|c|c|c|c|c|c|c|}
    \hline
        \multicolumn{3}{|c|}{ } &\multicolumn{4}{c|}{Number of events assigned to} & \multicolumn{2}{c|}{ }    \\
        \hline
        Extraction &	\makecell{Number of \\candidate \\events}	& \makecell{Number \\ fit to \\ $^{71}$Ge}	& \makecell{$^{51}$Cr source \\production}	& \makecell{Solar $\nu$ \\production}	& Carryover	& \makecell{$^{71}$Ge \\Production \\ decay rate  \\ (atoms/day)} & $Nw^2$	& Probability (\%) \\
        \hline
        Inner 9 Ex.	& 913	& 553.9	& 538.9	& 5.3	& 9.7	& 57.7$^{+3.2}_{-3.1}$	& 	0.093	& 26\\
        Outer 9 Ex. 	& 889	& 599.7	& 564.5	& 28.5	& 6.6	& 59.8$^{+3.1}_{-3.2}$	&	0.058	& 59\\
        \hline
    \end{tabular}
    \label{tab:results_without_first}
\end{table*}

\section{$^{71}$G\lowercase{e} Calibration} \label{app:ge71}

\subsection{Correction for Energy Calibration}

To verify the extrapolation for the K and L peak regions from the $^{55}$Fe calibration peak, counters were filled with $^{71}$Ge for additional analysis. 
Positions and resolutions of the peaks from extrapolation were compared to the true $^{71}$Ge peaks.
This analysis also enables to select events with real $T_N$ limits and verify the accepted value of 96\% acceptance. 

Table~\ref{tab:counter_calibration} summarizes the positions and the resolutions of the $^{55}$Fe calibration and the $^{71}$Ge K and L peaks. 
The values predicted from the calibration peak, taking into account the counter non-linearity factors, are presented with the true location of the X-ray peaks. 
Means and standard deviations of the two types of counter systems: sys2z~(new) and sys3~(old), are also presented for each peak. 
These values are used as correction factors for the event selection.

\subsection{Rise time window $T_N$ Selection Test}

The $^{71}$Ge calibration data is also used as a test to the $T_N$ limits event selection windows. 
The cutoff values, determined as the ratio of $T_N$ selected events to all events without selection, is expected to be close to 96\%.
The values presented in Table~\ref{tab:tn_test} agrees with 96\% within uncertainty. 
Hence the $T_N$ event selection efficiency is verified.

\section{Results without the First Data Points}

\begin{figure}[t]
    \centering
    \includegraphics[width=\columnwidth]{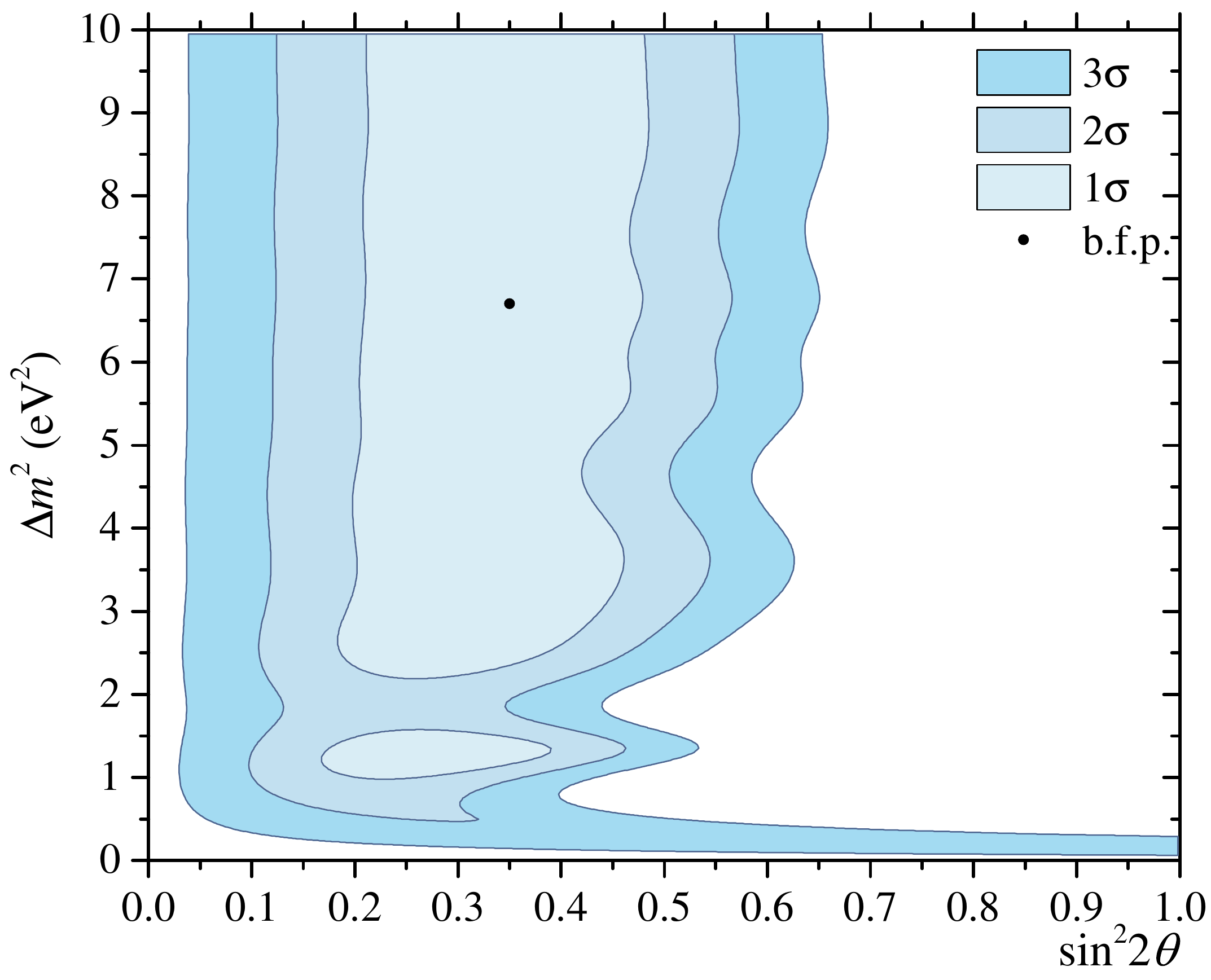}
    \includegraphics[width=\columnwidth]{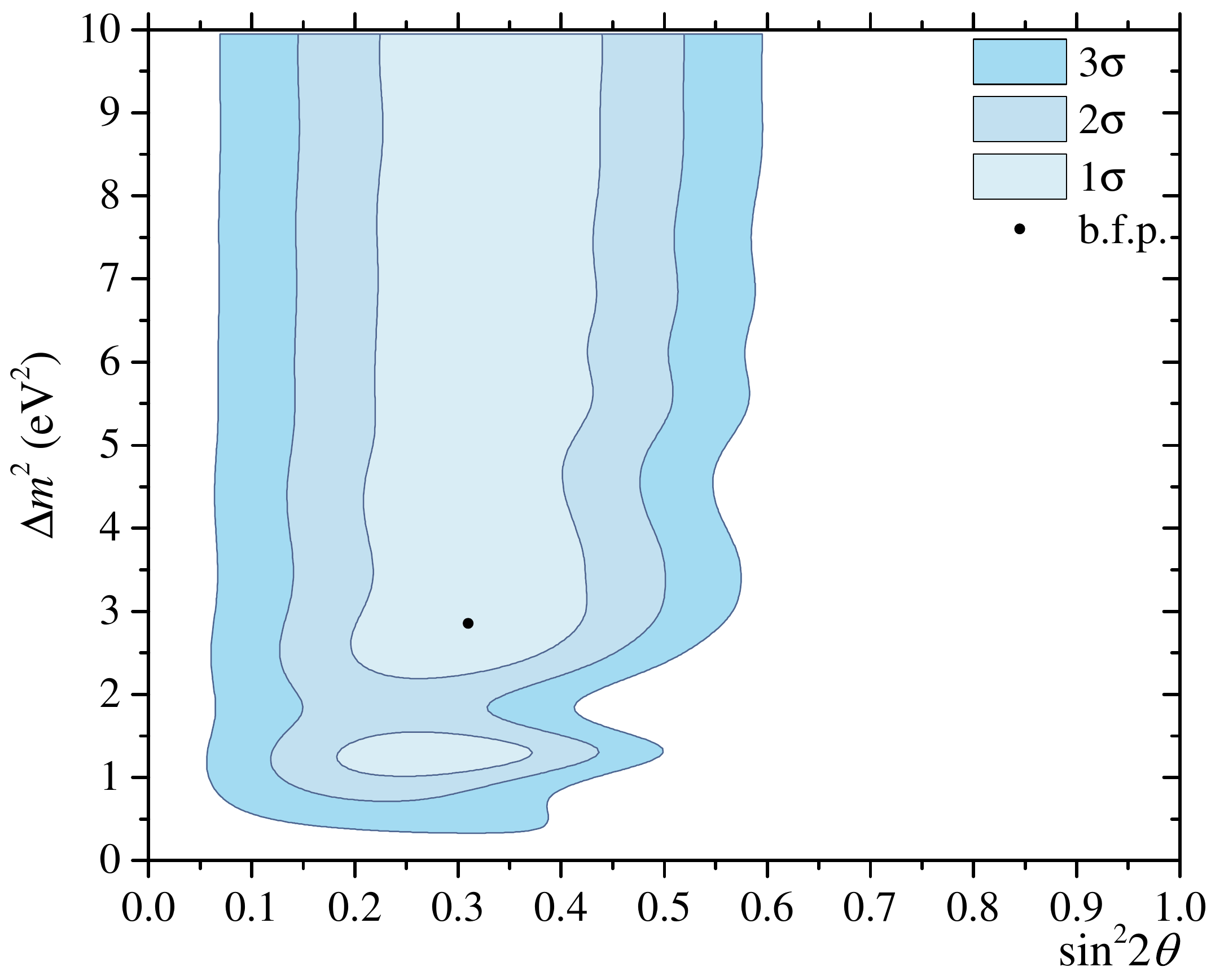}
    \caption{\textbf{Top:} Allowed regions for two BEST results without the first extractions. The best-fit point is sin$^22\theta$=0.35, $\Delta m^2$ =6.7~eV$^{2}$ and is indicated by a point. \textbf{Bottom:} Allowed regions for two GALLEX, two SAGE and two BEST results without the first extractions. The best-fit point is sin$^22\theta$=0.31, $\Delta m^2$  = 2.85~eV$^{2}$ and is indicated by a point.}
    \label{fig:wo_Cr1_allowed}
\end{figure}

In Fig.~\ref{fig:production_rate}, it seems that the most significant contributions to the deficit for both the inner and the outer volumes originate from the first extractions. 
To examine the effect of the first extraction, we carried out an analysis leaving out the first extraction. 
For the 9 runs of each target without the first extractions, the combined best fit rates are:

\begin{equation}
\begin{split}
    r_\mathrm{In-meas.} &=57.7\pm 3.2~\mathrm{atoms}^{71}\mathrm{Ge/day}~, \\
    r_\mathrm{Out-meas.} &= 59.8\pm 3.2~\mathrm{atoms}^{71}\mathrm{Ge/day}~. 
\end{split}
\end{equation}

\noindent
These are 3\% and 6\% increase in rates, respectively. 
The statistical uncertainty with 68\% confidence increases by 19\% in the two volumes.
The quadratic combination of all systematic uncertainties is -2.7/+2.9\%. 
The values are summarized in Table~\ref{tab:results_without_first}.

The measured production rates in the K and L peaks, including both statistical and systematic uncertainties combined in quadrature  are

\begin{equation}
\begin{split}
    r_\mathrm{In-meas.} &= 57.7 \pm3.2(\mathrm{stat.}) ^{+1.7}_{-1.6}(\mathrm{syst.}) \\
                        &= 57.7^{+3.6}_{-3.6}~\mathrm{atoms}^{71}\mathrm{Ge/day}~, \\
    r_\mathrm{Out-meas.} &= 59.8\pm3.2(\mathrm{stat.}) ^{+1.7}_{-1.6}(\mathrm{syst.}) \\
                         &= 59.8^{+3.6}_{-3.6}~\mathrm{atoms}^{71}\mathrm{Ge/day}~. 
\end{split}
\end{equation}

The ratio of measured to predicted production rates are

\begin{equation}
\begin{split}
    R&_\mathrm{In} = \frac{57.7^{+3.6}_{-3.6}}{69.4^{+2.5}_{-2.0}} = 0.83^{+0.06}_{-0.06}~,\\
    R&_\mathrm{Out} = \frac{59.8^{+3.6}_{-3.6}}{72.6^{+2.6}_{-2.1}} = 0.82^{+0.06}_{-0.06}~.
\end{split}
\end{equation}

\noindent
These are 2.9$\sigma$ and 3.1$\sigma$ less than unity.
The ratio of the outer to the inner result is $0.99\pm0.08$ and there is no difference in the capture rates in the two zones.
In Fig.~\ref{fig:wo_Cr1_allowed}, the allowed region for the BEST result and for all gallium anomaly experiments are illustrated.

\section{$^{51}$Cr Branching Ratio Uncertainty}

The calorimetric heat measurement relies on the branching ratio of the 320~keV $^{51}$Cr emission to normalize to activity.
If the branching value is in error, so would be the source strength.

\begin{table}[thp]
    \centering
    \caption{A summary of the branching ratio of the 320~keV emission from $^{51}$Cr.}
    \begin{tabular}{|l|l|l|}
    \hline
    Branching ratio & Reference & Method \\
    \hline
         0.1030(19) & \cite{FISHER1984121} & Ge(Li)\\
         0.0990(8) & \cite{KONSTANTINOV1994200} & NaI \\
         0.1008(11) & \cite{KONSTANTINOV1994200} & HPGe \\
         0.099(1) & \cite{YALCIN200563} & HPGe (Beta-gamma coincidence) \\
         0.0987(3) & \cite{MOREIRA2010596} & Si(Li) with fixed activity \\
     \hline
    \end{tabular}
    \label{tab:branching}
\end{table}

Table~\ref{tab:branching} summarizes various measurements of the branching ratio to the excited state. The branching ratios are claimed to be known to a precision $\sim$0.1\%, and vary within themselves by 4\%. 
This is much smaller than the observed gallium anomaly and is not enough to explain our results.

\newpage
\bibliography{BESTextended}

\end{document}